\documentclass[pra,twocolumn,showpacs,superscriptaddress,floatfix, nofootinbib,reprint, amsmath,amssymb,aps,longbibliography]{revtex4-2}

\usepackage{graphicx}
\usepackage{dcolumn}
\usepackage{bm}
\usepackage{braket}
\usepackage{float, quantikz, adjustbox, graphicx}
\usepackage[caption=false]{subfig}
\usepackage{bbold}

\usepackage{tabularray}

\usepackage[T1]{fontenc}

\tikzset{
operator/.append style={fill=white!20},
}
\tikzset{U1/.style={
    minimum width=1cm,
    minimum height=1cm,
}}

\tikzset{U1last/.style={
    minimum width=1cm,
    minimum height=1cm
}}

\tikzset{U1lastdag/.style={
    minimum width=1cm,
    minimum height=1cm,
}}

\tikzset{U2/.style={
    minimum width=1cm,
    minimum height=1cm,
}}

\begin{document}

\title{Scalable platform for qudit-based quantum computing using polar molecules}

\author{Soleh Kh. Muminov}
\affiliation{Russian Quantum Center, Skolkovo, Moscow 121205, Russia}
\affiliation{Moscow Institute of Physics and Technology, Institutsky lane 9, Dolgoprudny, Moscow region,
141700, Russia}
\author{Evgeniy O. Kiktenko}%
\affiliation{Russian Quantum Center, Skolkovo, Moscow 121205, Russia}
\affiliation{Department of Mathematical Methods for Quantum Technologies,
Steklov Mathematical Institute of Russian Academy of Sciences, Gubkina St.
8, Moscow 119991}
\author{Anastasiia S. Nikolaeva}
\affiliation{Russian Quantum Center, Skolkovo, Moscow 121205, Russia}
\affiliation{National University of Science and Technology ``MISIS”,  Moscow 119049, Russia}%
\author{Denis A. Drozhzhin}%
\affiliation{Russian Quantum Center, Skolkovo, Moscow 121205, Russia}
\affiliation{National University of Science and Technology ``MISIS”,  Moscow 119049, Russia}
\author{Sergey I. Matveenko}%
\affiliation{Russian Quantum Center, Skolkovo, Moscow 121205, Russia}
\affiliation{L. D. Landau Institute for Theoretical Physics, Chernogolovka 142432, Russia}
\author{Aleksey K. Fedorov}%
\affiliation{Russian Quantum Center, Skolkovo, Moscow 121205, Russia}
\affiliation{National University of Science and Technology ``MISIS”,  Moscow 119049, Russia}%
\author{Georgy V. Shlyapnikov}%
\affiliation{Russian Quantum Center, Skolkovo, Moscow 121205, Russia}
\affiliation{Universit\'e Paris-Saclay, CNRS, LPTMS, 91405 Orsay, France}
\affiliation{Van der Waals-Zeeman Institute, Institute of Physics, University of Amsterdam,
Science Park 904, 1098 XH Amsterdam, The Netherlands}

\newcommand{\beginsupplement}{%
        \setcounter{table}{0}
        \renewcommand{\thetable}{S\arabic{table}}%
        \setcounter{figure}{0}
        \renewcommand{\thefigure}{S\arabic{figure}}%
     }

\begin{abstract}
We propose a scalable qudit-based quantum processor using rotational states of polar molecules.
Previously, molecular internal states were used to enlarge Hilbert space, whereas our approach uses optical tweezer arrays to achieve scalable architectures with exponential state-space growth without increasing qudit dimensionality $d$.
Entangling gates are implemented by adiabatically bringing traps together to activate dipole-dipole interactions.
We develop encoding schemes mapping single qubits into qudits with $2\leq d\leq5$ and pairs of qubits into $d=4,5$ qudits, enabling universal set of quantum gates.
Additional levels in $d=3$ and $d=5$ qudits simplify multiqubit gate decompositions.
We analyze experimental parameters for SrF and NaCs molecules.
This approach provides a promising route to scalable quantum information processing with multilevel systems using existing experimental platforms.
\end{abstract}

\maketitle

\section{Introduction}\label{sec:intro}

Recent progress in quantum information technologies has led to significant interest in the use of qudits (multilevel quantum systems) in view of their potential to enhance the efficiency of the execution of quantum algorithms~\cite{muthukrishnan2000multivalued,Bartlett2002Quantum,zilic2007scaling,Parasa2011Quantum} and error correction schemes~\cite{campbell2012magic,duclos2013kitaev,campbell2014enhanced,anwar2014fast,andrist2015error,watson2015qudit,krishna2019towards}. 
Various quantum algorithms can be implemented in a more efficient manner with qudits (for review, see Ref.~\cite{kiktenko2025colloquium}) due to the fact that, first, a single qudit can be presented as a set of qubits~\cite{Clark2004,Kiktenko2015,Kiktenko20152} and, second, additional levels of qudits can be used as ancillas for simplifying the realization of multiqubit gates, such as the Tofolli gate~\cite{Ralph2007,Wallraff2012,Goss2023Toffoli,kiktenko2020scalable,Nikolaeva2025}. The combination of these approaches may lead to a substantial reduction in circuit depth~\cite{kiktenko2025colloquium}, which is of special interest for the current generation of noisy intermediate-scale quantum devices (NISQ)~\cite{Aspuru-Guzik2022}, where both the number of qubits and the fidelities of the gates are limited.

\begin{figure}
    \centering
    \includegraphics[width=0.9\linewidth]{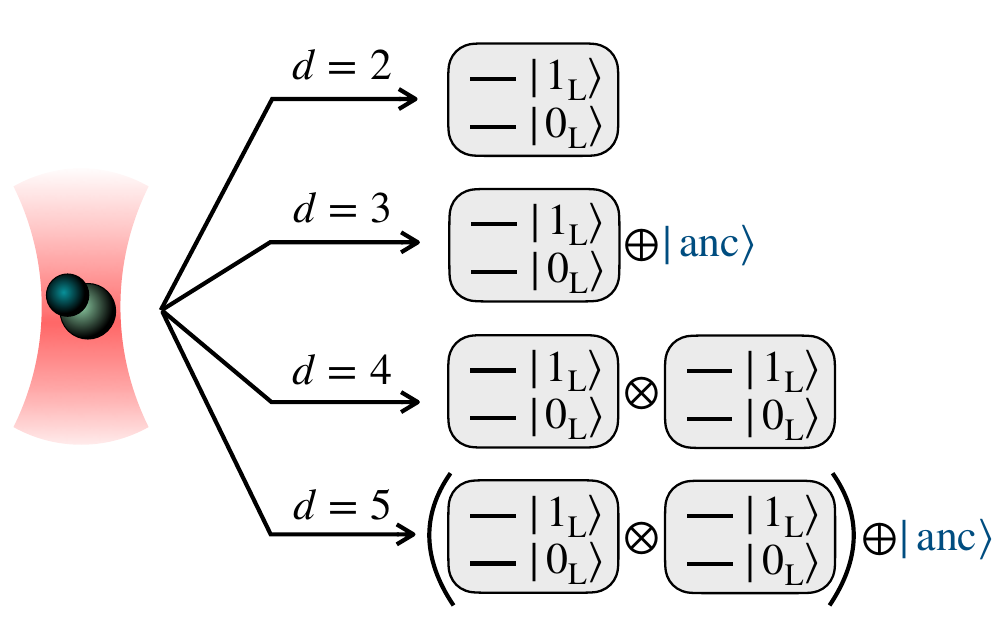}
    \caption{Encoding schemes for qubits within the internal state space of a single molecular dipole, as explored in this work. The presented schemes utilize various dimensions (d) of the Hilbert space.  The $d$=2 scheme employs a standard qubit encoding within the molecular states.  A three-level system ($d$=3) is proposed, where two levels encode a qubit, and the third one serves as an ancilla for simplifying multi-qubit gate decomposition. Furthermore, a four-level encoding ($d$=4) enables the encoding of two qubits within a single molecule.  Finally, a five-level scheme ($d$=5) is introduced, accommodating two qubits and an ancilla level, mirroring the functionality of the $d$=3 scheme.}
    \label{fig:general}
\end{figure}

Among various physical systems studied for qudit implementations~\cite{morvan2021qutrit,goss2022high,sung2021multi,ringbauer2022universal,zalivako2024towards,fu2022experimental,zhou2024robust, Morello2024nature, Zache2022}, polar molecules have emerged as a particularly promising platform~\cite{sawant2020ultracold}. Recent progress in experiments with cold polar molecules~\cite{ni2008high,moses2015creation,wall2015realizing,manmana2012topological,anderegg2018laser,deiglmayr2008formation,ospelkaus2010quantum,deiglmayr2008formation,ni2010dipolar,de2011controlling,yan2013observation,hazzard2014many,carr2009cold,andre2006coherent,kaufman2021quantum} has opened opportunities to use them in quantum computing~\cite{demille2002quantum,gregory2021robust,picard2025entanglement,ni2018dipolar,bao2023dipolar,caldwell2020long,ruttley2025long} and metrology~\cite{mitra2022quantum,hudson2006cold,zelevinsky2008precision,demille2008enhanced,sauer2006probing,vutha2008search}. A particular feature of such systems is the possibility of controlling and manipulating the occupation of energy levels, in particular by using strong, tunable dipole-dipole interactions~\cite{demille2002quantum,sawant2020ultracold,picard2025entanglement}. 
The rotational, vibrational, and electronic degrees of freedom inherent to these systems allow precise quantum state manipulations by external electromagnetic fields. Notably, rotational states present exceptional advantages for qudit encoding, as they exhibit long coherence times due to isolation from environmental decoherence channels. Furthermore, permanent electric dipole moments characteristic of heteronuclear molecules provide a mechanism for establishing strong intermolecular couplings, which is a critical requirement for scalable quantum system architectures~\cite{demille2002quantum,hughes2020robust,herrera2014infrared}. Experimental progress in laser cooling techniques~\cite{shuman2010laser} and optical confinement of ultracold molecular ensembles~\cite{Anderegg2019optical,ruttley2023formation}, has significantly advanced the prospects for practical realization of dipole-mediated quantum information processing.

Rotational states of ultracold polar molecules have long been recognized as a natural and powerful resource for quantum information processing. Early theoretical proposals demonstrated that microwave-addressable rotational levels, combined with strong and tunable dipole-dipole interactions, enable universal quantum computation with molecular qubits~\cite{demille2002quantum,yelin2006schemes}. Subsequent work established rotational manifolds as effective spin degrees of freedom for quantum simulation and quantum logic in optical lattices and tweezer arrays~\cite{micheli2006toolbox,gorshkov2011tunable}. Historically, however, experimental implementations have often encoded long-lived storage qubits in hyperfine sublevels within a single rotational manifold, motivated by their extended coherence times and reduced sensitivity to electric-field noise~\cite{park2017second,gregory2021robust,gregory2024second,ospelkaus2010controlling}, while rotational excitations were primarily employed as auxiliary states to mediate entangling interactions~\cite{demille2002quantum,yelin2006schemes}. This strategy was largely driven by early observations of limited rotational coherence~\cite{blackmore2018ultracold}.

Recent advances in molecular control -- including improved electric-field stability, optical lattice and tweezer trapping~\cite{Anderegg2019optical,bao2023dipolar}, coherent microwave manipulation~\cite{will2016coherent,ji2020microwave}, and dynamical decoupling -- have substantially extended the coherence times of rotational states. On the experimental side, coherent control and dipolar interactions between rotational levels have been demonstrated in lattice- and tweezer-trapped polar molecules, including dipolar spin-exchange dynamics and entangling two-qubit gates~\cite{yan2013observation,ni2018dipolar}. Most notably, the recent demonstration of second-scale coherence across multiple rotational levels~\cite{hepworth2025long} firmly establishes rotational manifolds themselves as a viable computational basis. These developments provide the essential experimental foundation for qudit encodings based directly on rotational states and clarify the relative advantages and challenges of rotational- versus hyperfine-based encoding schemes~\cite{blackmore2018ultracold}, thereby motivating the scalable rotational qudit architectures developed in the present work.

\begin{figure*}
    \centering
    \includegraphics[width=1.0\linewidth]{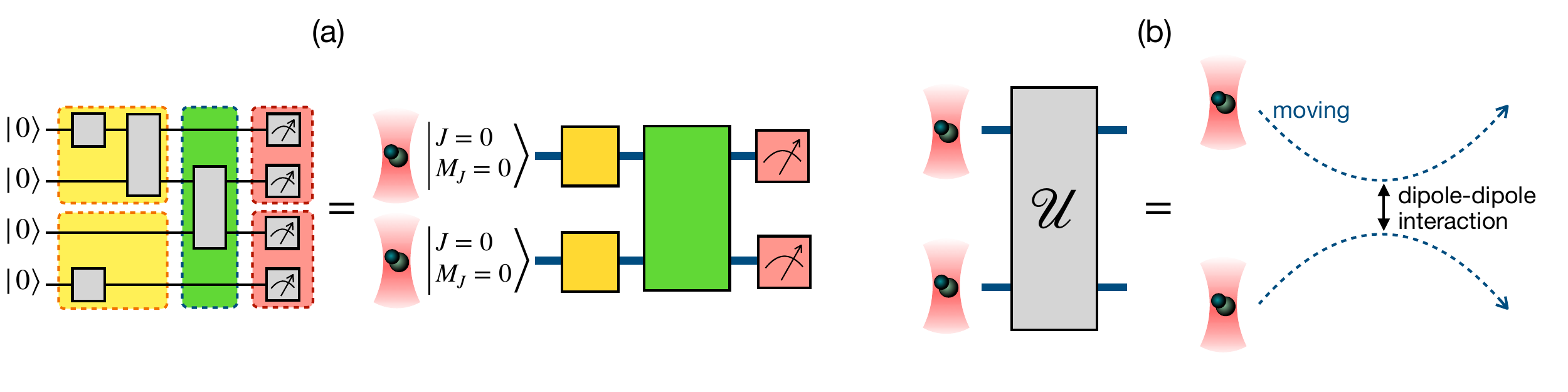}
    \caption{(a) Schematic illustration of implementing a qubit-based quantum circuit using dipolar qudits, where each qudit encodes two qubits ($d=4$). Single-qudit gates correspond to operations on qubits within the same molecule, while entangling gates between qubits in different qudits are realized via interactions between distinct molecules.  
    (b) Conceptual depiction of the entangling gate between molecules, achieved by bringing two optical tweezers (and their respective dipoles) to close proximity to enable dipole-dipole interaction and then followed by their separation.}
    \label{fig:entangling}
\end{figure*}

In this work, we develop a scheme for universal qubit-based quantum computing using rotational levels of individual ultracold polar molecules confined in optical tweezer arrays, treating each molecule as a qubit or qudit of dimension $d = 3, 4, 5$ (see Fig.~\ref{fig:general}).
The computational space is encoded in the subset of rotational levels: (i) encoding qubits in two-level subspaces; (ii) utilizing three-level subspaces to include an auxiliary level that simplifies multiqubit gate decompositions such as Toffoli gates; (iii) employing four-level subspaces to encode two qubits per molecule; and (iv) exploiting five-level subspaces to encode two qubits plus an auxiliary level for streamlined implementation of complex multiqubit operations. 
Although higher-dimensional encodings introduce additional experimental complexity, they offer enhanced flexibility and efficiency in the execution of quantum algorithms.

The key result of our work is the construction of a universal gate set including single-qubit rotations and two-qubit entangling gates acting on qubits encoded within the same or different molecules [Fig.~\ref{fig:entangling}(a)]. Unlike Refs.~\cite{gregory2021robust,yelin2006schemes,wei2011entanglement,ni2018dipolar,picard2025entanglement}, which consider $J=0$ states in a magnetic field, we use states with different $J$ and $M_J$. For the molecules studied, excited rotational states with $J>0$ have radiative lifetimes much longer than operation and coherence times. Entangling gates are realized by bringing molecular tweezers close to each other to enhance dipole-dipole interactions and then separating the molecules [Fig.~\ref{fig:entangling}(b)]. We numerically validate this scheme for SrF and NaCs. Our scalability does not rely on increasing qudit dimension. Early qudit proposals emphasized encoding multiple qubits in a single molecule using large internal manifolds \cite{sawant2020ultracold,gregory2021robust,zhang2022creation,Luis2020Gd2,Hernándezantón2024}, motivated by limited experimental scalability. Recent progress in large-scale optical tweezer arrays and coherent dipolar control \cite{bao2023dipolar,zhang2022optical,ruttley2024enhanced} enables scalable architectures based on increasing molecule number instead of internal dimensionality. Based on this, we combine fixed-dimensional rotational qudits with scalable tweezer platforms to provide an experimentally realistic route to large-scale quantum information processing. As a result, the Hilbert space grows with particle number (without increasing $d$), enabling flexible implementation of complex algorithms. Combining multilevel rotational encoding, long coherence times, and a scalable universal gate architecture with molecular entangling operations makes this platform a strong candidate for scalable quantum computing.



The paper is organized as follows.
In Sec.~\ref{sec:eqs} we present a comprehensive examination of rotational-state manifolds in diatomic polar molecules as a foundation for qudit encoding. This section further elucidates the mechanisms by which dipole-dipole interactions may be used to generate entanglement between adjacent qudit systems.
In Sec.~\ref{sec:QIP}, we develop encodings for qudits with dimensions $d = 2, 3, 4$, and 5, and construct universal gate sets for each encoding.
In Sec.~\ref{sec:Decoherence} we estimate the fundamental limitations to coherence times, and, thus, the fidelity of the operations. Finally, we summarize our results and conclude in Sec.~\ref{sec:Concl}.

\section{Operating with dipole-based qudits} \label{sec:eqs}

We consider a system of identical heteronuclear diatomic molecules interacting with each other via dipole-dipole interaction (DDI). In the adiabatic approximation, neglecting spin effects, molecular energy levels can be classified as electronic, vibrational, and rotational levels~\cite{landau2013quantum}. 
Throughout our work, we assume that the molecules are in their electronic and vibrational ground states and in their low rotational states.
We denote the rotational states of a single molecule by $\ket{J, M_{J}}$, where $J=0,1,\ldots$ is the rotational moment and $M_J=-J,\ldots,J$ is its projection.
The rotational part of the Hamiltonian for a single molecule can be expressed as:
\begin{equation} \label{eq:1q-Hamiltonian}
    \begin{aligned}
        \hat H_{\rm rot}&=\sum_{J}
    E_{\rm rot}(J)\sum_{M_J}\ket{J,M_J}\bra{J, M_J},\\
    E_{\rm rot}(J)&=BJ(J+1),
    \end{aligned}
\end{equation}
where $B$ is the rotational constant.

In this work, we consider finite $d$-element subsets of the set of all possible rotational states $\{|J,M_J\rangle\}$ as computational states for a qudit.
Single-qudit operations of the form
\begin{equation} \label{eq:1q-gate}
    \hat R^{{Q}{Q}'}_{x(y)}(\theta)=\exp\left(-i\hat\sigma_{x(y)}^{{Q}{Q}'}\frac{\theta}{2}\right),
\end{equation}
can be induced by resonant Rabi oscillations. Here ${Q}=(J,M_J)$, ${Q}'=(J',M_J')$ are two distinct pairs of quantum numbers, $\theta$ is the Bloch sphere rotation angle, and:
\begin{equation}
    \begin{aligned}
        \hat\sigma_{x}^{{Q}{Q}'}&:=\ket{J, M_J}\bra{J', M_J'}+{\rm h.c.},\\
    \hat\sigma_{y}^{{Q}{Q}'}&:=-i\ket{J, M_J}\bra{J', M_J'} +{\rm h.c.}    
    \end{aligned}
\end{equation}
We note that a sufficient condition for realizing the gate $\hat R^{QQ'}_{x(y)}(\theta)$ for arbitrary $Q$ and $Q'$ is the availability of a set of individually addressable transitions between selected pairs of levels forming a connected graph in the qudit state space.
Although each addressable transition is driven by its own characteristic frequency, the total number of distinct frequencies required remains fixed.
It has been shown that an arbitrary unitary operation on a $d$-level qudit can be implemented using at most $d(d-1)/2$ two-level addressable operations~\cite{Brennen2005,drozhzhin2025,nikolaeva2024efficient}.
Therefore, direct controllability of all possible level-to-level transitions is not required.


Next, we consider an implementation of the entangling operation between dipolar qudits.
Consider a system of two molecules with time-dependent Hamiltonian
\begin{equation}
    \hat{H}_{\rm sys}(t) = \hat{H}_1 + \hat{H}_2 + \hat{V} [R(t)],
\end{equation}
where $\hat{H}_1=\hat{H}_{\rm rot}\otimes\hat{\mathbb{1}}$ and $\hat{H}_2=\hat{\mathbb{1}}\otimes \hat{H}_{\rm rot}$ correspond to the internal Hamiltonians~\eqref{eq:1q-Hamiltonian} of the first and second molecules, respectively (here $\hat{\mathbb{1}}$ stands for the identity operator), $\hat{V}$ represents the dipole-dipole interaction, and $R(t)$ is the intermolecular separation, which is treated as a time-dependent parameter. 
Initially, at time $t=0$, the intermolecular separation is assumed to be large, and therefore the interaction energy $\hat{V}$ can be neglected. 
To entangle the molecules, they are slowly brought together to a minimum separation and then moved again to a larger distance over time $t=\tau$. During this process, the DDI alters the population of the molecular states, thereby applying a two-qudit gate.

The time evolution of the system state $\ket{\psi(t)}$ is governed by the time-dependent Schrödinger equation.
Consider the projection of $\ket{\psi(t)}$ onto the basis states of the noninteracting molecules:
\begin{equation}
    c_{\bf Q}(t):=\left(\bra{J_1,M_{J_1}}\otimes\bra{J_2,M_{J_2}}\right)e^{-iE_{\bf Q}t/\hbar}\ket{\psi(t)},
\end{equation}
where ${\bf Q}=(J_1,M_{J_1};J_2,M_{J_2})$ is a multiindex of quantum numbers for two molecules and $E_{\bf Q}=E_{\rm rot}(J_1)+E_{\rm rot}(J_2)$.
The time-dependent Schr{\"o}dinger equation can then be expressed as~\cite{landau2013quantum}:
\begin{equation}\label{eq:SysDiffEq}
    \frac{\partial c_{\bf Q}(t)}{\partial t} = \frac{1}{i \hbar} \sum_{{\bf Q}'} c_{{\bf Q}'}(t) V_{{\bf Q}{\bf Q}'}(t) e^{i(E_{{\bf Q}} - E_{{\bf Q}'})t/\hbar},
\end{equation}
where $V_{{\bf Q}{\bf Q}'}(t)$ is the matrix element of $\hat{V}[R(t)]$.

The slow approach and subsequent separation of molecules, governed by $R(t)$ for $t\in[0,\tau]$, allows us to neglect transitions between quantum states of different energies, in accordance with the adiabatic theorem. 
Consequently, the vibrational states remain unchanged, and transitions occur solely between rotational states which have the same energies.



Using the spherical harmonic addition theorem, the DDI operator can be expressed as:
\begin{multline}
    \hat{V}(\theta_1,\phi_1;\theta_2,\phi_2;R) \\= \frac{d^2}{R^3} \left(\sin\theta_1\sin\theta_2 \cos(\phi_1-\phi_2) - 2\cos \theta_1 \cos \theta_2 \right),
\end{multline}
where $d$ represents the magnitude of the dipole moment, $R$ is the distance between the molecules, and the polar (azimuthal) angles $\theta_i$ ($\phi_i$) specify the orientation of the $i$-th molecule. 
The matrix element of this operator takes the form:
\begin{widetext}
    \begin{multline}
        V_{{\bf Q}{\bf Q}'}(t)=\bra{J_1,M_{J_1}}\otimes\bra{J_2,M_{J_2}}\hat{V}[R(t)]
        \ket{J_1',M_{J_1}'}\otimes\ket{ J_2',M_{J_2}'}\\
        = \frac{d^2}{R^3(t)} \sqrt{\frac{(2J_1+1)(2J_2+1)}{(2J_1'+1)(2J_2'+1)}} C_{1,0;J_1,0}^{J_1',0} C_{1,0;J_2,0}^{J_2',0} \left(\sum_{k=-1}^1 C_{1,k;J_1,M_{J_1}}^{J_1',M_{J_1}'} C_{1,-k;J_2,M_{J_2}}^{J_2',M_{J_2}'} + C_{1,0;J_1,M_{J_1}}^{J_1',M_{J_1}'} C_{1,0;J_2,M_{J_2}}^{J_2',M_{J_2}'} \right),
    \end{multline}
\end{widetext}
where $C_{J_1,M_{J_1};J_2,M_{J_2}}^{J,M{J}'}$ are the Clebsch-Gordan coefficients. It can be seen that there are the following selection rules: $\Delta J_1 = \pm 1$, $\Delta J_2 = \pm 1$, $\Delta M_{J_1} = 0, \pm 1$, $\Delta M_{J_2} = 0, \mp 1$.

To determine the evolution operator matrix, we solve the system of Eqs.~\eqref{eq:SysDiffEq} for a given time-dependent function $R(t)$ numerically. 
In our calculations, we chose three forms of $R(t)$:
\begin{equation} \label{eq:R-of-t}
    \begin{aligned}
        R(t) &= \begin{cases}
            \alpha \cos\left(\frac{\pi (x+2) t}{\tau} \right) + \beta, t \le \frac{\tau}{x+2} \\
            \beta - \alpha, \frac{\tau}{x+2} < t \le \frac{x+1}{x+2}\tau\\
            \alpha \cos\left(\frac{\pi(x+2)t}{\tau} - \pi x\right) + \beta, \frac{x+1}{x+2}\tau < t \le \tau\\
            \alpha + \beta, t > \tau
        \end{cases}\\
    \end{aligned}
\end{equation}
\begin{equation} \label{eq:R2-of-t}
    \begin{aligned}
        R(t) &= \begin{cases}
            -\frac{2\alpha x t}{\tau} + \beta + \alpha, t \le \tau/x\\
            \beta-\alpha, \tau/x < t \le \tau(1-1/x)\\
            \frac{2\alpha x t}{\tau} + \beta + \alpha(1-2x), \tau(1-1/x) < t \le \tau\\
            \alpha + \beta, t > \tau
        \end{cases}\\
    \end{aligned}
\end{equation}
\begin{equation} \label{eq:R3-of-t}
    \begin{aligned}
        R(t) &= \begin{cases}
            \alpha \left(1- \tanh\left(\frac{2 x t}{\tau}-3\right) \right) + \beta - \alpha, t \le \frac{\tau}{2}\\
            \alpha \left(1- \tanh\left(\frac{-2 x t}{\tau} + 2x-3 \right) \right) + \beta - \alpha, \frac{\tau}{2} < t \le \tau\\
            \alpha(1-\tanh(-3)) + \beta - \alpha, t > \tau.
        \end{cases}
    \end{aligned}
\end{equation}
where $\alpha$ and $\beta$ are parameters such that $\beta + \alpha$ represents the initial (maximum) intermolecular separation and $\beta - \alpha$ represents the minimum intermolecular separation, and $\tau$, $x$ are tunable parameters.

If the function $R(t)$ varies sufficiently slowly, then in accordance with Eq. (\ref{eq:SysDiffEq}), transitions are allowed only between states with the same energy in a two-molecule system (the vibrational and electronic states of the molecules also remain unchanged). For this condition to hold, the characteristic timescale $\tau$ must significantly exceed $\hbar / B$. At the same time, $\tau$ should be much smaller than the coherence time $\tau_c$ of the rotational states of the molecules. Thus, the time $\tau$ should satisfy the inequality:
\begin{equation}\label{eq:tauInequality}
    \tau_c \gg \tau \gg \hbar/B.
\end{equation}

The time $\tau$ satisfying the inequality (\ref{eq:tauInequality}) we select such that the probability of transition from the state $\ket{1,0}\otimes\ket{0,0}$ to $\ket{0,0}\otimes\ket{1,0}$ for $R(t)$ from Eq. (\ref{eq:R-of-t}) is close to $1$. It turns out that for such $\tau$ the resulting evolution operator has the same form as an iSWAP-type gate acting in the subspace with the following basis states ${\cal Q}\otimes {\cal Q}$:
\begin{equation}
    {\cal Q}=\{\ket{0,0},\ket{1,0},\ket{3,-3},\ket{3,0},\ket{3,3}\},  
\end{equation}
and the same for $R(t)$ from Eqs. (\ref{eq:R2-of-t})-(\ref{eq:R3-of-t}).
We note that the present approach does not impose any fundamental restriction on the accessible qudit dimensionality $d$. 
However, increasing the number of internal levels with the aim of encoding more qubits into a single qudit introduces a practical overhead: implementing two-qubit operations between qubits stored in different qudits generally requires a larger number of entangling gates. 
A concrete instance of this issue is encountered in Sec.~\ref{sec:ququarts}. A more detailed discussion of scaling challenges can be found in Ref.~\cite{kiktenko2025colloquium}.

The target entangling gate, which acts in the subspace formed by the basis states ${\cal Q}\otimes {\cal Q}$, has the form:
\begin{equation} \label{eq:2q-gate}
    \begin{aligned}
        {\cal U}\ket{1,0}\otimes\ket{0,0}&= i\ket{0,0}\otimes\ket{1,0}\\
        {\cal U}\ket{0,0}\otimes\ket{1,0}&= i\ket{1,0}\otimes\ket{0,0}\\
        {\cal U}\ket{J,M_J}\otimes\ket{J',M_J'}&= \ket{J,M_J}\otimes\ket{J',M_J'},
    \end{aligned}
\end{equation}
where $\ket{J,M_J}\ket{J',M_J'}$ are taken from the set of $23$ states ${\cal Q}\otimes {\cal Q}$ $\backslash$ $\{\ket{0,0}\otimes\ket{1,0}, \ket{1,0}\otimes\ket{0,0}\}$.
One can think about the resulting ${\cal U}$ as a direct sum of a two-qubit iSWAP gate in the subspace ${\rm Span}(\{\ket{0,0},\ket{1,0}\}^{\otimes 2})$ and identity in ${\rm Span}(\{\ket{3,-3},\ket{3,0},\ket{3,3}\}^{\otimes 2})$.
We note that the specific choice of levels in ${\cal Q}$ that excludes states with $J=2$ leads to an interaction ${\cal U}$ of this particularly convenient form for implementing quantum gates. States with $J=2$ were included in the simulations but were excluded from the final protocol because most of their magnetic sublevels exhibit significant population dynamics due to their coupling structure and energy spacing, leaving only a single stable state. This is insufficient for defining a robust qudit basis, which requires two stable states with nearly constant populations. In contrast, the $J=3$ manifold provides multiple sublevels with negligible population evolution during gate operations and is therefore better suited for the proposed qudit scheme.
The target cost function for maximization is taken in the form:
\begin{equation}
    {\cal F}_{\rm iSWAP}=\frac{1}{25}|{\rm Tr}({\cal U}^{\rm phys}_{\cal Q}{\cal U}^\dagger)|,
\end{equation}
where ${\cal U}^{\rm phys}_{\cal Q}$ is a projection of the solution of Eqs.~\eqref{eq:SysDiffEq} for given $R(t)$ on the 25-level subspace of ${\cal Q}\otimes {\cal Q}$ .
We also calculate the fidelity of the two-qubit identity gate:
\begin{equation}
    {\cal F}_{\rm id}=\frac{1}{25}|{\rm Tr}{\cal I}^{\rm phys}_{\cal Q}|,
\end{equation}
where ${\cal I}^{\rm phys}_{\cal Q}$ is a projection of the evolution operator for molecules remaining at rest at the distance $R(t)=\overline{r}$ during $t\in[0,\tau]$.


 Numerical solutions of the system of differential equations (\ref{eq:SysDiffEq}) are obtained for SrF molecule in the X$^2\Sigma^+$ electronic state and NaCs molecule in the X$^1\Sigma^+$ state. The values of the physical parameters are taken from ~\cite{ernst1985electric,hao2019high,haimberger2009formation}. The molecules are considered to be in a 100 $\mu$K deep potential well. The results for the $R(t)$ dependence, given by Eq. (\ref{eq:R-of-t}) are represented in Table~\ref{tab:NumericalOptimization}. The parameters $\tau$ for $R(t)$ given by Eqs. (\ref{eq:R2-of-t}) and (\ref{eq:R3-of-t}) differ from $\tau$ given by Eq. (\ref{eq:R-of-t}) by a numerical coefficient smaller than 3.
In all cases, the fidelities ${\cal F}_{\rm iSWAP}$ and ${\cal F}_{\rm id}$ are found to be close to 1, within the limits of numerical accuracy.

Also we quantified leakage probability outside the computational subspace $\cal{Q}$. For an arbitrary normalized state $\ket{\psi} \in \cal{Q}$, the probability to remain within the computational subspace after the gate operation is given by $\bra{\psi} ({\cal U}^{\rm phys}_{\cal Q})^\dagger {\cal U}^{\rm phys}_{\cal Q} \ket{\psi}|$. The worst-case leakage probability is therefore determined by the minimal eigenvalue of the positive-semidefinite matrix $({\cal U}^{\rm phys}_{\cal Q})^\dagger {\cal U}^{\rm phys}_{\cal Q}$:
\begin{equation}
    P_{\rm leak}^{\rm worst} = 1 - \lambda_{\rm min},
\end{equation}
where $\lambda_{\rm min}$ -- denotes the minimal eigenvalue of $({\cal U}^{\rm phys}_{\cal Q})^\dagger {\cal U}^{\rm phys}_{\cal Q}$. $P_{\rm leak}^{\rm worst}$ provides a strict upper bound on population transfer out of $\cal{Q}$ for any logical input state.

For the optimized trajectories $R(t)$ and molecular parameters considered in this work, we find $P_{\rm leak}^{\rm worst} = 3.6 \cdot 10^{-6}$. This result demonstrates that leakage outside the computational subspace is strongly suppressed by the combined effect of adiabatic evolution and dipole-dipole selection rules. Consequently, the entangling gate dynamics remains effectively confined to $\cal{Q}$, and leakage errors are negligible on the scale relevant for the gate fidelities reported above.

When an optical tweezer moves along a prescribed trajectory $R(t)$ the molecule experiences an inertial force $F(t)= - m \ddot{R}(t)$. For a particle initially prepared in the motional ground state of a harmonic trap with frequency $\omega$, the final motional excitation depends on the Fourier components of the driving at the trap frequency. In particular, the probability of exciting vibrational quanta is suppressed when the motion of the trap center is sufficiently slow compared to the intrinsic timescale. A convenient adiabaticity parameter is: \begin{equation}
    \eta=\frac{a_{\rm max}}{\omega^2 x_{ho}},
\end{equation}
where $a_{\rm max} = \max |\ddot{R}(t)|$, $x_{ho} = \sqrt{\hbar/(m \omega)}$. The transport is adiabatic and excitations of the center-of-mass motion are negligible provided that $\eta \ll 1$, a condition that is satisfied for the parameters listed in Table~\ref{tab:NumericalOptimization}.

The dipole moment $d$ of the molecule and the minimum distance $r_0$ at which molecules can approach each other are the crucial parameters. At present, for the schemes using a single laser, we have $r_0$ close to 1 $\mu$m or larger~\cite{florshaim2024spatial}. For more complicated schemes using two lasers, one can think of significantly smaller $r_0$~\cite{caldwell2020enhancing,caldwell2021general}, and even with a single laser, $r_0$ can be smaller than 1 $\mu$m~\cite{sWill}. For $r_0 \approx 1 \mu$m, the time $\tau$ is still much smaller than the coherence time $\tau_c$ ($\tau_c$ related to the ac-Stark effect is of the order of tens of msec or even larger~\cite{burchesky2021rotational,bause2020tune,blackmore2020controlling}). The time $\tau$ is also smaller than the coherence time of the rotational states of the molecules with $J > 0$ related to the dipole-dipole interaction at the maximum intermolecular separation. To obtain a higher fidelity of two qubit gates, it is reasonable to consider the molecule with a larger dipole moment. The situation is much better for the minimum distance $r_0 \approx$ 0.6 $\mu$m, although even in this case an increase of the dipole moment $d$ is desirable.

\begin{table}
\centering
\caption{Parameters of the $R(t)$ dependence obtained after numerical optimization. SrF and NaCs were chosen as target molecules.}
\label{tab:NumericalOptimization}
\begin{tblr}{
  column{even} = {c},
  column{3} = {c},
  column{5} = {c},
  cell{1}{2} = {c=2}{},
  cell{1}{4} = {c=2}{},
  cell{2}{2} = {c=2}{},
  cell{2}{4} = {c=2}{},
  cell{3}{2} = {c=2}{},
  cell{3}{4} = {c=2}{},
  cell{4}{2} = {c=4}{},
  hlines,
  vlines,
}
                                                                                           & SrF  &     & NaCs &     \\
$d$, D                                                                                   & 3.5  &     & 4.6  &     \\
$B$, cm$^{-1}$                                                                    & 0.25 &     & 0.06 &     \\
initial distance ($\beta+\alpha = \overline{r}$), $\mu$m & 10   &     &      &     \\
minimum distance ($\beta-\alpha = r_0$), $\mu$m & 0.6  & 1   & 0.6  & 1   \\
$\tau$, $\mu$sec                                      & 80   & 237 & 47   & 142 
\end{tblr}
\end{table}

\section{Quantum computing with dipolar qudits}\label{sec:QIP}

\begin{figure*}
    \centering
    \includegraphics[width=0.7\linewidth]{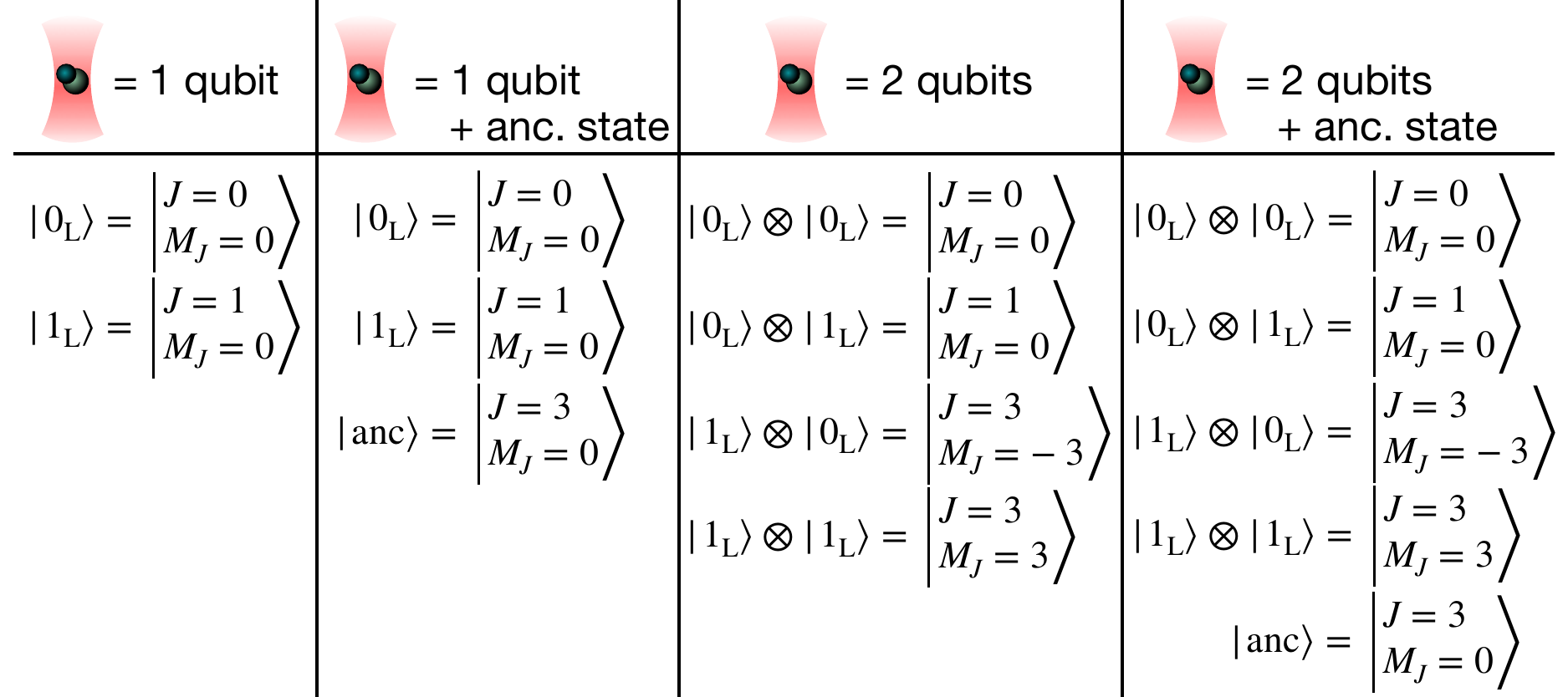}
    \caption{A summary of the schemes for encoding a single or two qubits in a single dipole with $d=2,3,4,$ and 5 (columns from left to right), as discussed in the main text.}
    \label{fig:encodings}
\end{figure*}

In this section, we explore the potential use of the described system for running qubit-based quantum circuits.  
We propose several methods for encoding qubits in dipolar qudits and construct universal gate sets that are applicable to these encodings.
We assume that the following operations can be performed: (i) single-qudit gates of the form~\eqref{eq:1q-gate} for states within the set ${\cal Q}$, (ii) entangling gate of the form~\eqref{eq:2q-gate}, and (iii) measurement of a single qudit in the ${\cal Q}$ basis.
We further assume that the initial state of a qudit register is $\ket{0,0}\otimes\ldots\otimes\ket{0,0}$, and that any leakage from ${\cal Q}$ can be neglected for the purposes of this consideration.
Below, we consider encoding schemes in
$d$-dimensional subspaces of ${\rm Span}(\mathcal{Q})$ for $d = 2, 3, 4,$ and 5, as illustrated in Fig.~\ref{fig:encodings}.

\subsection{Molecular dipole as qubit}
We start with a straightforward encoding of a qubit in a 2-dimensional subspace:
\begin{equation}
        \ket{0,0} \leftrightarrow \ket{0_{\rm L}},\quad
        \ket{1,0} \leftrightarrow \ket{1_{\rm L}}.
\end{equation}
(Hereinafter, we use subindex $L$ to denote logical states of encoded qubits.)

Then single-qubit gates can be directly realized via $R^{0,0;1,0}_{x(y)}(\theta)$, while ${\cal U}$ can be used as a two-qubit iSWAP gate:
\begin{equation} \label{eq:qubit_iSWAP}
    \begin{aligned}
        {\cal U} \ket{0_{\rm L}}\otimes \ket{0_{\rm L}} &= \ket{0_{\rm L}}\otimes \ket{0_{\rm L}}, \\
        {\cal U} \ket{0_{\rm L}} \otimes\ket{1_{\rm L}} &=  i\ket{1_{\rm L}}\otimes \ket{0_{\rm L}}, \\
        {\cal U} \ket{1_{\rm L}}\otimes \ket{0_{\rm L}} &=  i\ket{0_{\rm L}}\otimes \ket{1_{\rm L}}, \\
        {\cal U} \ket{1_{\rm L}}\otimes \ket{1_{\rm L}} &= \ket{1_{\rm L}}\otimes \ket{1_{\rm L}}.
    \end{aligned}
\end{equation}
The inverse of the iSWAP gate with $i$ replaced by $-i$ can be created by using a pair of Pauli $Z$ gates $Z\otimes Z={\rm diag}(1,-1,-1,1)$ before (or after) the iSWAP gate. 
Single-qubit $Z$ can be obtained by applying $R_{x}^{1_{\rm L}A}(2\pi)$, where $A$ denotes a level in ${\cal Q}$ other than $\ket{0,0}$ and $\ket{1,0}$.
We also note that the controlled NOT (CNOT) gate can be realized by using two iSWAP gates with single-qubit operations~\cite{williams2011quantum}.

\subsection{Molecular dipole as qubit and ancillary level}

The second approach extends the previous one by adding an ancillary logical level $\ket{{\rm anc}}$:
\begin{equation}
    \ket{0,0} \leftrightarrow \ket{0_{\rm L}}, \quad 
    \ket{1,0} \leftrightarrow \ket{1_{\rm L}}, \quad
    \ket{3,0} \leftrightarrow \ket{{\rm anc}}.
\end{equation}
As in the previous case, the universal gate set can be constructed on the basis of single-qubit gates and iSWAP gate~\eqref{eq:qubit_iSWAP}.
At the same time, by surrounding ${\cal U}$ with single-qudit gates ${\sf P}$ and ${\sf P}^\dagger$, where
\begin{equation}
     {\sf P} := R^{0_{\rm L},{\rm anc}}_y(-\pi)~R^{0_{\rm L},1_{\rm L}}_y(\pi) \\
 \end{equation}
provides the following transformation:
\begin{equation}
   {\sf P} \ket{0_{\sf L}} = \ket{1_{\sf L}},\quad 
   {\sf P} \ket{1_{\sf L}} = \ket{\sf anc}, \quad
   {\sf P} \ket{\sf anc} = \ket{0_{\sf L}},
\end{equation}
and we can synthesize the operation in the following manner:
\begin{equation}
    {\sf iSWAP}^{0_{\rm L},{\rm anc}} := (\mathbb{1}\otimes{\sf P})~{\cal U}~(\mathbb{1}\otimes{\sf P}^\dagger)
\end{equation}
which acts as follows:
\begin{equation}
    \begin{aligned}
        &{\sf iSWAP}^{0_{\rm L},{\rm anc}}\ket{1_{\rm L}}\otimes\ket{1_{\rm L}}=i\ket{0_{\rm L}}\otimes\ket{{\rm anc}},\\
        &{\sf iSWAP}^{0_{\rm L},{\rm anc}} \ket{0_{\rm L}}\otimes\ket{{\rm anc}}=i\ket{1_{\rm L}}\otimes\ket{1_{\rm L}},\\
        &{\sf iSWAP}^{0_{\rm L},{\rm anc}} \ket{x_{\rm L}}\otimes\ket{y_{\rm L}}=\ket{x_{\rm L}}\otimes\ket{y_{\rm L}}
    \end{aligned}
\end{equation}
for $\ket{x_{\rm L}}\ket{y_{\rm L}}\in\{\ket{0},\ket{1},\ket{\rm anc}\}^{\otimes 2}\backslash\{\ket{1_{\rm L}}\ket{1_{\rm L}},\ket{0_{\rm L}}\ket{{\rm anc}}\}$.
This gate can be used to decompose a generalized $N$-qubit controlled-phase gate ${\sf C}^{N-1}{\sf Z}$:
\begin{equation}\label{eq-cnz}
\begin{gathered}
    {\sf C}^{N-1}{\sf Z}\ket{1_{\rm L}}\otimes\ldots\otimes\ket{1_{\rm L}}=-\ket{1_{\rm L}}\otimes\ldots\otimes\ket{1_{\rm L}},\\
    {\sf C}^{N-1}{\sf Z}\ket{x^1_{\rm L}}\otimes\ldots\otimes\ket{x^{N-1}_{\rm L}}=\ket{x^1_{\rm L}}\otimes\ldots\otimes\ket{x^{N-1}_{\rm L}},
\end{gathered}
\end{equation}
where $\prod_{i=1}^N x_i\neq1$ using $2N-2$ application of ${\cal U}$ as shown in~\cite{nikolaeva2022decomposing}.

\begin{figure}
    \centering
\includegraphics[width=0.98\linewidth]{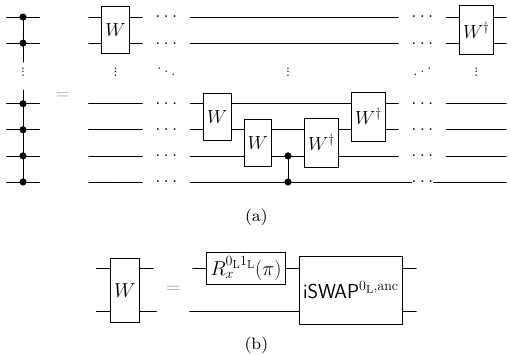}
    \caption{(a) ${\sf C}^{N-1}{\sf Z}$ decomposition with $2N-4$ $W$ operations and two ${\sf iSWAP}^{0_{\rm L},{\rm anc}}_1$ gates (for the implementation of ${\sf CZ}$ in the center of the scheme), when dipole is considered as a qutrit: qubit with ancillary state.
    (b) Definition of the operation $W$ with ${\sf iSWAP}$ gate and single-qudit rotation. }
    \label{fig:cnz-qutrits}
\end{figure}

The idea of the decomposition shown in  Fig.~\ref{fig:cnz-qutrits}(a) is as follows. 
First, we note that the multiqubit gate in Eq.~\eqref{eq-cnz} applies the phase factor $-1$ to all $N$ involved qubits if and only if they are in the state $\ket{1_{\rm L}}^{\otimes L}$; otherwise, the state of the qubits remains the same.
To decompose ${\sf C}^{N-1}{\sf Z}$, we introduce an operation $W$ that acts on $N-2$ pairs of qudits:
\begin{equation}\label{eq-w}
    \begin{aligned}
        &W\ket{0_{\rm L}}\otimes\ket{0_{\rm L}}=\ket{1_{\rm L}}\otimes\ket{0_{\rm L}},\\
        &W\ket{0_{\rm L}}\otimes\ket{1_{\rm L}}= i\ket{0_{\rm L}}\otimes\ket{{\rm anc}},\\
        &W\ket{1_{\rm L}}\otimes\ket{0_{\rm L}}=\ket{0_{\rm L}}\otimes\ket{0_{\rm L}},\\
        &W\ket{1_{\rm L}}\otimes\ket{1_{\rm L}}=\ket{0_{\rm L}}\otimes\ket{1_{\rm L}},
    \end{aligned}
\end{equation}
and whose construction is shown in Fig.~\ref{fig:cnz-qutrits}(b).
The operator $W$ is designed such that the bottom qubit remains in the state $|1_{\rm L}\rangle$ only when the inputs are in the state $\ket{1_{\rm L}}\otimes \ket{1_{\rm L}}$.
Therefore, after the sequence of $W$ operations in Fig.~\ref{fig:cnz-qutrits}(a), if the lower qubit comes to the state $\ket{1_{\rm L}}$, it implies that each qubit affected by the sequence was initially in the state $\ket{1_{\rm L}}$.
In this case, the application of a central controlled phase gate will add a $-1$ phase if and only if all qubits were in the $\ket{1_{\rm L}}$ initial state.
Finally, a series of uncomputation operations $W^\dagger$ is applied to remove populations of all ancillary states $\ket{{\rm anc}}$.

The use of ${\sf iSWAP}$ gate as a basic entangling operation makes the decomposition in Fig.~\ref{fig:cnz-qutrits}(a) adaptive for arbitrary coupling map between dipoles (see~\cite{nikolaeva2022decomposing} for more details).
In particular, the depth $\mathcal{O}(\log N)$ of the decomposition circuit is achieved for a binary-tree coupling map. 
We also note that the multiqubit Toffoli gate can be obtained from the ${\sf C}^{N-1}{\sf Z}$ gate by applying a pair of single-qudit gates to the target before and after ${\sf C}^{N-1}{\sf Z}$.

\subsection{Molecular dipole as two qubits}\label{sec:ququarts}
Next we consider encoding of a pair of qubits in a single $4$-level dipole (ququart) according to the following mapping between qubit and qudit states:
\begin{equation}
    \begin{aligned}
        &\ket{0,0}\leftrightarrow \ket{0_{\rm L}0_{\rm L}},\\
        &\ket{1,0}\leftrightarrow \ket{0_{\rm L}1_{\rm L}},\\
        &\ket{3,-3}\leftrightarrow\ket{1_{\rm L}0_{\rm L}},\\
        &\ket{3,3}\leftrightarrow \ket{1_{\rm L}1_{\rm L}}.\\
    \end{aligned}\label{eq:4mapping}
\end{equation}

All single-qubit gates and two-qubit gates for qubits embedded in the same qudit can be realized via local operations~\cite{nikolaeva2024efficient}.
In particular, a single qubit gate $u$ can be considered as a tensor product of the form $u\otimes\mathbb{1}$ or  $\mathbb{1}\otimes u$, where $\mathbb{1}$ is the $2\times 2$ identity acting on a neighboring, unaffected qubit that resides in the same qudit.
These tensor products can then be straightforwardly decomposed into two-level operations. 
The entangling iSWAP gate acting on two qubits of the same qudit can be realized as $R_x^{(0_{\rm L}1_{\rm L}),(1_{\rm L}0_{\rm L})}(-\pi/2)$.

\begin{figure}[H]
    \centering
\hfill\hfill\subfloat[]{\resizebox{.37\linewidth}{!}{
\includegraphics[width=\linewidth]{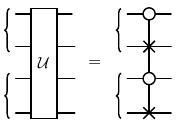}
}}\hfill
\subfloat[]{\resizebox{.5\linewidth}{!}{
\includegraphics[width=0.9\linewidth]{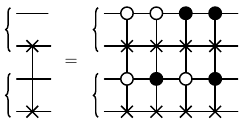}
}}
    \caption{(a) The scheme of the controlled iSWAP gate.
    (b) The synthesis of the CNOT gate between qubits from distinct qudits.
    Vertical braces denote qubits that belong to the same qudit. We note that the $\times$ symbols used here represent iSWAP rather than a standard SWAP operation, as is typically the case.}
    \label{fig:CiSWAP}
\end{figure}

Regarding entangling gates between qubits residing in different qudits, these can be constructed as follows.  
According to the mapping, the ${\cal U}$ operation acts as
\begin{equation}
    \begin{aligned}
        {\cal U}\ket{0_{\rm L}0_{\rm L}}\otimes\ket{0_{\rm L}1_{\rm L}}&=i\ket{0_{\rm L}1_{\rm L}}\otimes\ket{0_{\rm L}0_{\rm L}},\\
        {\cal U}\ket{0_{\rm L}1_{\rm L}}\otimes\ket{0_{\rm L}0_{\rm L}}&=i\ket{0_{\rm L}0_{\rm L}}\otimes\ket{0_{\rm L}1_{\rm L}},\\
    \end{aligned}
\end{equation}
and acts as the identity operator on the remaining $4^2-2=14$ levels of joint space of four qubits. 
One can see that this is a $\ket{0_{\rm L}}^{\otimes 2}$-controlled iSWAP gate, where the iSWAP operation is applied to the second and fourth qubits when the first and the third qubits are in the all-zero state  [see Fig.~\ref{fig:CiSWAP}(a)].
The realization of a standard iSWAP gate between qubits from distinct qudits is illustrated in Fig.~\ref{fig:CiSWAP}(b).
In this construction, the control pattern can be modified by applying single-qubit inversion gates as needed.

\subsection{Molecular dipole as two qubits with joint ancillary level}
A single 5-level dipole (ququint) can be considered as a pair of qubits with a joint ancillary level according to the following mapping of the states:
\begin{equation}
    \begin{aligned}
        &\ket{0,0}\leftrightarrow \ket{0_{\rm L}0_{\rm L}},\\
        &\ket{1,0}\leftrightarrow \ket{0_{\rm L}1_{\rm L}},\\
        &\ket{3,-3}\leftrightarrow\ket{1_{\rm L}0_{\rm L}},\\
        &\ket{3,3}\leftrightarrow \ket{1_{\rm L}1_{\rm L}},\\
        &\ket{3,0}\leftrightarrow \ket{\rm anc}.
    \end{aligned}\label{eq:5mapping}
\end{equation}
We note that mapping (\ref{eq:5mapping})~in a 4-level subspace, which corresponds to the space of two qubits, is equal to mapping (\ref{eq:4mapping}).
Therefore, single-qubit gates and two-qubit gates, for all possible embeddings of qubits in qudit space, are implemented as it is described in section \ref{sec:ququarts}.

However, due to the presence of joint ancillary level we can efficiently decompose multiqubit gates~\cite{nikolaeva2024efficient}.
In particular, generalized 4-qubit controlled phase gate ${\sf C}^3{\sf Z}$ can be implemented with two successive applications of ${\cal U}$ operations and single-ququint gates as it is shown in Fig.~\ref{fig:cccz}.


\newcommand{\gry}[1]{\gate[wires=2, style={inner ysep=-0.08cm}]{#1}}

\begin{figure*}
    \centering
\resizebox{.65\linewidth}{!}{\begin{quantikz}[column sep=0.2cm, font=\large]
&\ctrl{1} & \midstick[4, brackets=none]{~$=$~} &\gry{R_y^{0_{\rm L}1_{\rm L},{\text{anc}}}(-\pi)}&\gry{R_y^{0_{\rm L}0_{\rm L},1_{\rm L}1_{\rm L}}(-\pi)}&\gate[wires=4]{\mathcal{U}^2}&\gry{R_y^{0_{\rm L}0_{\rm L},1_{\rm L}1_{\rm L}}(\pi)}&\gry{R_y^{0_{\rm L}1_{\rm L},{\text{anc}}}(\pi)}&\\[-0.8cm]
&\ctrl{1} &&&&&&&\\[-0.3cm]
&\ctrl{1} &&\gry{R_y^{0_{\rm L}1_{\rm L},{\text{anc}}}(-\pi)}&\gry{R_y^{0_{\rm L}1_{\rm L},1_{\rm L}1_{\rm L}}(-\pi)}&&\gry{R_y^{0_{\rm L}1_{\rm L},1_{\rm L}1_{\rm L}}(\pi)}&\gry{R_y^{0_{\rm L}1_{\rm L},{\text{anc}}}(\pi)}&\\[-0.8cm]
&\ctrl{0} &&&&&&&\\
\end{quantikz}}
    \caption{${\sf C}^3{\sf Z}$ decomposition with $\mathcal{U}$ operation and single-ququint rotations.}
    \label{fig:cccz}
\end{figure*}
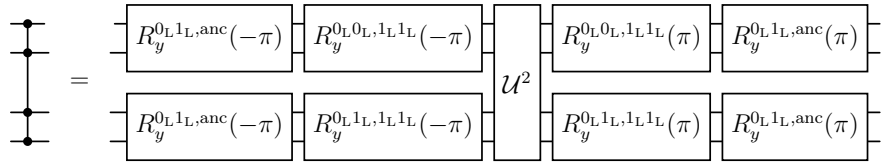

\section{Sources of decoherence and gate errors}\label{sec:Decoherence}

In this section we discuss the physical processes that primarily limit coherence times (and therefore the achievable fidelity) in systems of ultracold polar molecules trapped in optical tweezers. To provide concrete estimates, we assume 1064 nm tweezers with a numerical aperture NA $\sim$ 0.7, a trap depth of the order of $\sim$ 100 $\mu$K, and a molecular temperature $\sim$ 2 $\mu$K.

At the minimum separation $0.6 \mu$m several effects arise from interference between the optical fields of neighboring tweezers, leading to modifications of the trapping potentials. The most relevant effects are photon scattering, molecule tunneling between adjacent tweezers and collisional losses. The tunneling induced by the tensor part of the Stark shift operator can be strongly suppressed by choosing orthogonal polarizations of the tweezers as it is demonstrated in Supplementary materials for Ref.~\cite{caldwell2020enhancing}. Collisional loss rates were also  estimated following~\cite{caldwell2020enhancing} (Supplementary, Eq. (S20)). Photon scattering becomes the dominant fundamental limitation on coherence at sufficiently large trap depths. However, using Eqs. (12) and (13) from~\cite{grimm1999opticaldipoletrapsneutral} we estimate scattering rate of the order of $\mathrm{\Gamma}_{sc} \sim 0.1-0.3$ s$^{-1}$ for NaCs and $\mathrm{\Gamma}_{sc} \sim 0.5$ s$^{-1}$ for SrF. These rates imply upper-bound coherence times of approximately $\tau_c \sim 3-4$s for NaCs molecule and $\tau_c \sim 2$s for SrF molecule. Furthermore, in the far-off-resonant regime most scattering events are Rayleigh-type, whereas Raman scattering, which changes the internal (rotational) state, is suppressed by one to two orders of magnitude relative to the total scattering rate~\cite{cline1994spin,caldwell2020sideband,kotochigova2024rotational}. Our estimates include Rayleigh and rotational Raman scattering induced by the trapping light but neglect hyperfine-changing Raman transitions. For large detunings, these are suppressed due to small hyperfine splittings and interference effects~\cite{kotochigova2024rotational,grimm1999opticaldipoletrapsneutral,cline1994spin}. Thus, photon scattering mainly preserves the rotational manifold, although population transfer to undetectable hyperfine states may act as an effective loss channel depending on the detection scheme. Coherence times are therefore determined by the total photon scattering rate.


The dominant source of dephasing for interaction-based gates is the variation of the dipolar interaction strength $J$. Assuming quasi-static fluctuations between experimental realizations (as in Ref.~\cite{bergonzoni2025iswap}), we identify two principal mechanisms. First, thermal motion of the molecules  --  particularly along the axial direction of the tweezers  --  causes fluctuations in the intermolecular separation. To estimate this effect we can use Eq.~(56) from Ref.~\cite{bergonzoni2025iswap}, with the difference that $\Delta J_{\rm mot} / J$ depends on time. Under our assumptions it gives us the maximum value of approximately $0.116$ for $\Delta J_{\rm mot} / J$ at $R = r_0$ for NaCs (and almost the same for SrF) and $\Delta J_{\rm mot} / J \approx 4 \cdot 10^{-4}$ at $R = \overline{r}$ for both molecules. Decreasing temperature to 1 $\mu$K leads to a decrease in the maximum value of $\Delta J_{\rm mot} / J$ to approximately $0.062$ at $R = r_0$ for NaCs (and to $\Delta J_{\rm mot} / J \approx 0.065$ for SrF). The second mechanism arises from motion–rotation coupling. Even in the adiabatic transport regime ($\eta \ll 1$), where excitations of the center-of-mass motion are strongly suppressed, residual motional effects give rise to fluctuations in the effective dipolar coupling $J$. In particular, state-dependent trap displacements associated with motion–rotation coupling lead to variations of $J$, which contribute to gate imperfections and are quantified below by the relative fluctuation $\Delta J_{\rm mot-rot}/J$. Using an estimated relative trap-center shift of $\zeta \approx 1.5$nm between the $\ket{0}$ and $\ket{1}$ states, we obtain $\Delta J_{\rm mot-rot}/J \approx 0.0015$ for NaCs molecule and $\Delta J_{\rm mot-rot} / J \approx 0.0021$ for SrF. These values indicate that fluctuations of $J$ due to the motion-rotation coupling can be omitted compared to (thermal) motional fluctuations.

While technical noise sources (intensity fluctuations, beam pointing, electric field noise, polarization drifts) may further reduce coherence in practice, the above estimates characterize the fundamental physical limits under idealized control conditions. Although, these effects are experiment-specific and can be mitigated with active stabilization; therefore we do not include them explicitly in our theoretical error budget.

A detailed analysis of the influence of the fine and hyperfine structure on the proposed scheme is provided in the Supplemental Material \cite{SM}.

\section{Conclusion} \label{sec:Concl}

We have developed a scalable qudit platform based on diatomic polar molecules, which is among the leading platforms for quantum technologies. Our approach uses the rotational levels of the molecules with qudit dimensions $d = 2, 3, 4$ and 5. A universal gate set has been proposed, and entangling gates between qudits are created relying on molecules moving in optical traps and exhibiting coherent couplings due to dipole-dipole interactions. Numerical simulations for SrF and NaCs molecules have demonstrated the feasibility of this approach.
Specifically, it has been shown that the gate times can be shorter than the coherence times of the rotational levels, which opens the way to realizing quantum algorithms with such systems.  This scalable architecture offers a promising path towards robust and efficient quantum computation using ultracold polar molecules.
Advantages of the proposed approach in experiments can be demonstrated by reducing the number of operations required to realize quantum algorithms, such as Grover's search and Bernstein-Vazirani algorithm.

\section*{Acknowledgements}
We are grateful to A. M. Rey, S. Will, A. V. Akimov and S. S. Straupe for fruitful discussions.
The work of E.O.K. on designing the universal gate set for dipole qubits was supported by the state assignment of the Ministry of Science and Higher Education of the Russian Federation (Section IIIA).
The work of A.K.F. is supported by the Priority 2030 program at the NUST ``MISIS'' under the project K1-2022-027 (developing gate set for qubit pairs, encoded in dipole qudits; Section IIIC).
The work of A.S.N. and D.A.D. was supported by RSF Grant No.~24-71-00084 (developing realization of multicontrolled-phase gate on dipole qudits; Sections IIIB and IIID).

\nocite{apsrev42Control}
\bibliographystyle{apsrev4-2}
\bibliography{apssamp}

@CONTROL{apsrev42Control,
  author="08",
  editor="1",
  pages="0",
  title="0",
  year="1"
}

@PREAMBLE{
 "\providecommand{\noopsort}[1]{}" 
 # "\providecommand{\singleletter}[1]{#1}%" 
}

@article{Brennen2005,
   title={Criteria for exact qudit universality},
   volume={71},
   ISSN={1094-1622},
   url={http://dx.doi.org/10.1103/PhysRevA.71.052318},
   DOI={10.1103/physreva.71.052318},
   number={5},
   journal={Phys. Rev. A},
   publisher={American Physical Society (APS)},
   author={Brennen, Gavin and O’Leary, Dianne and Bullock, Stephen},
   year={2005},
   month=may }

@misc{drozhzhin2025,
      title={Transition-Aware Decomposition of Single-Qudit Gates}, 
      author={Denis A. Drozhzhin and Evgeniy O. Kiktenko and Aleksey K. Fedorov and Anastasiia S. Nikolaeva},
      year={2025},
      eprint={2510.25561},
      archivePrefix={arXiv},
      primaryClass={quant-ph},
      url={https://arxiv.org/abs/2510.25561}, 
}

@article{Aspuru-Guzik2022,
  title = {Noisy intermediate-scale quantum algorithms},
  author = {Bharti, Kishor and Cervera-Lierta, Alba and Kyaw, Thi Ha and Haug, Tobias and Alperin-Lea, Sumner and Anand, Abhinav and Degroote, Matthias and Heimonen, Hermanni and Kottmann, Jakob S. and Menke, Tim and Mok, Wai-Keong and Sim, Sukin and Kwek, Leong-Chuan and Aspuru-Guzik, Al\'an},
  journal = {Rev. Mod. Phys.},
  volume = {94},
  issue = {1},
  pages = {015004},
  numpages = {69},
  year = {2022},
  month = {Feb},
  publisher = {American Physical Society},
  doi = {10.1103/RevModPhys.94.015004},
  url = {https://link.aps.org/doi/10.1103/RevModPhys.94.015004}
}

@article{Kiktenko2015,
  title = {Multilevel superconducting circuits as two-qubit systems: Operations, state preparation, and entropic inequalities},
  author = {Kiktenko, E. O. and Fedorov, A. K. and Man'ko, O. V. and Man'ko, V. I.},
  journal = {Phys. Rev. A},
  volume = {91},
  issue = {4},
  pages = {042312},
  numpages = {7},
  year = {2015},
  month = {Apr},
  publisher = {American Physical Society},
  doi = {10.1103/PhysRevA.91.042312},
  url = {https://link.aps.org/doi/10.1103/PhysRevA.91.042312}
}

@article{Clark2004,
  title = {Maximizing the Hilbert Space for a Finite Number of Distinguishable Quantum States},
  author = {Greentree, Andrew D. and Schirmer, S. G. and Green, F. and Hollenberg, Lloyd C. L. and Hamilton, A. R. and Clark, R. G.},
  journal = {Phys. Rev. Lett.},
  volume = {92},
  issue = {9},
  pages = {097901},
  numpages = {4},
  year = {2004},
  month = {Mar},
  publisher = {American Physical Society},
  doi = {10.1103/PhysRevLett.92.097901},
  url = {https://link.aps.org/doi/10.1103/PhysRevLett.92.097901}
}

@article{Kiktenko20152,
	author = {E.O. Kiktenko and A.K. Fedorov and A.A. Strakhov and V.I. Man'ko},
	doi = {https://doi.org/10.1016/j.physleta.2015.03.023},
	issn = {0375-9601},
	journal = {Phys. Lett. A},
	number = {22},
	pages = {1409-1413},
	title = {Single qudit realization of the Deutsch algorithm using superconducting many-level quantum circuits},
	url = {https://www.sciencedirect.com/science/article/pii/S0375960115002753},
	volume = {379},
	year = {2015}}

@article{Nikolaeva2025,
  title = {Scalable Improvement of the Generalized Toffoli Gate Realization Using Trapped-Ion-Based Qutrits},
  author = {Nikolaeva, Anastasiia S. and Zalivako, Ilia V. and Borisenko, Alexander S. and Semenin, Nikita V. and Galstyan, Kristina P. and Korolkov, Andrey E. and Kiktenko, Evgeniy O. and Khabarova, Ksenia Yu. and Semerikov, Ilya A. and Fedorov, Aleksey K. and Kolachevsky, Nikolay N.},
  journal = {Phys. Rev. Lett.},
  volume = {135},
  issue = {6},
  pages = {060601},
  numpages = {6},
  year = {2025},
  month = {Aug},
  publisher = {American Physical Society},
  doi = {10.1103/p1z9-6w93},
  url = {https://link.aps.org/doi/10.1103/p1z9-6w93}
}

@article{kiktenko2025colloquium,
  title={Colloquium: Qudits for decomposing multiqubit gates and realizing quantum algorithms},
  author={Kiktenko, Evgeniy O and Nikolaeva, Anastasiia S and Fedorov, Aleksey K},
  journal={Rev. Mod. Phys.},
  volume={97},
  number={2},
  pages={021003},
  year={2025},
  publisher={APS}
}

@article{kiktenko2020scalable,
  title={Scalable quantum computing with qudits on a graph},
  author={Kiktenko, Evgeniy O and Nikolaeva, Anastasiia S and Xu, Peng and Shlyapnikov, Georgy V and Fedorov, Arkady K},
  journal={Phys. Rev. A},
  volume={101},
  number={2},
  pages={022304},
  year={2020},
  publisher={APS}
}

@article{Goss2023Toffoli,
  title={Empowering a qudit-based quantum processor by traversing the dual bosonic ladder},
  author={Nguyen, Long B and Goss, Noah and Siva, Karthik and Kim, Yosep and Younis, Ed and Qing, Bingcheng and Hashim, Akel and Santiago, David I and Siddiqi, Irfan},
  journal={Nat. Commun.},
  volume={15},
  number={1},
  pages={7117},
  year={2024},
  publisher={Nature Publishing Group UK London}
}

@article{Ralph2007,
	author = {Ralph, T. C. and Resch, K. J. and Gilchrist, A.},
	doi = {10.1103/PhysRevA.75.022313},
	issue = {2},
	journal = {Phys. Rev. A},
	month = {Feb},
	numpages = {5},
	pages = {022313},
	publisher = {American Physical Society},
	title = {Efficient Toffoli gates using qudits},
	url = {https://link.aps.org/doi/10.1103/PhysRevA.75.022313},
	volume = {75},
	year = {2007}}

@article{Wallraff2012,
	author = {Fedorov, A. and Steffen, L. and Baur, M. and da Silva, M. P. and Wallraff, A.},
	date = {2012/01/01},
	doi = {10.1038/nature10713},
	id = {Fedorov2012},
	isbn = {1476-4687},
	journal = {Nature},
	number = {7380},
	pages = {170--172},
	title = {Implementation of a Toffoli gate with superconducting circuits},
	url = {https://doi.org/10.1038/nature10713},
	volume = {481},
	year = {2012}}

@misc{Hernándezantón2024,
      title={Optimal Control of Spin Qudits Subject to Decoherence Using Amplitude-and-Frequency-Constrained Pulses}, 
      author={Alonso Hernández-Antón and Fernando Luis and Alberto Castro},
      year={2024},
      eprint={2403.15785},
      archivePrefix={arXiv},
      primaryClass={quant-ph},
      url={https://arxiv.org/abs/2403.15785}, 
}

@article{Luis2020Gd2,
  title={A dissymmetric [Gd2] coordination molecular dimer hosting six addressable spin qubits},
  author={Luis, Fernando and Alonso, Pablo J and Roubeau, Olivier and Velasco, Ver{\'o}nica and Zueco, David and Aguil{\`a}, David and Mart{\'\i}nez, Jes{\'u}s I and Barrios, Leon{\'\i} A and Arom{\'\i}, Guillem},
  journal={Communications Chemistry},
  volume={3},
  number={1},
  pages={176},
  year={2020},
  publisher={Nature Publishing Group UK London}
}

@article{Zache2022,
	author = {Gonz\'alez-Cuadra, Daniel and Zache, Torsten V. and Carrasco, Jose and Kraus, Barbara and Zoller, Peter},
	doi = {10.1103/PhysRevLett.129.160501},
	issue = {16},
	journal = {Phys. Rev. Lett.},
	month = {Oct},
	numpages = {8},
	pages = {160501},
	publisher = {American Physical Society},
	title = {Hardware Efficient Quantum Simulation of Non-Abelian Gauge Theories with Qudits on Rydberg Platforms},
	url = {https://link.aps.org/doi/10.1103/PhysRevLett.129.160501},
	volume = {129},
	year = {2022}}

@article{Morello2024nature,
  title={Navigating the 16-dimensional Hilbert space of a high-spin donor qudit with electric and magnetic fields},
  author={Fern{\'a}ndez de Fuentes, Irene and Botzem, Tim and Johnson, Mark AI and Vaartjes, Arjen and Asaad, Serwan and Mourik, Vincent and Hudson, Fay E and Itoh, Kohei M and Johnson, Brett C and Jakob, Alexander M and others},
  journal={Nat. Commun.},
  volume={15},
  number={1},
  pages={1380},
  year={2024},
  publisher={Nature Publishing Group UK London}
}

@article{demille2002quantum,
  title={Quantum computation with trapped polar molecules},
  author={DeMille, David},
  journal={Phys. Rev. Lett.},
  volume={88},
  number={6},
  pages={067901},
  year={2002},
  publisher={APS}
}

@article{morvan2021qutrit,
  title={Qutrit randomized benchmarking},
  author={Morvan, Alexis and Ramasesh, Vinay V and Blok, Machiel S and Kreikebaum, John Mark and O’Brien, K and Chen, Larry and Mitchell, Bradley K and Naik, Ravi K and Santiago, David I and Siddiqi, Irfan},
  journal={Phys. Rev. Lett.},
  volume={126},
  number={21},
  pages={210504},
  year={2021},
  publisher={APS}
}

@article{goss2022high,
  title={High-fidelity qutrit entangling gates for superconducting circuits},
  author={Goss, Noah and Morvan, Alexis and Marinelli, Brian and Mitchell, Bradley K and Nguyen, Long B and Naik, Ravi K and Chen, Larry and J{\"u}nger, Christian and Kreikebaum, John Mark and Santiago, David I and others},
  journal={Nat. Commun.},
  volume={13},
  number={1},
  pages={7481},
  year={2022},
  publisher={Nature Publishing Group UK London}
}

@article{sung2021multi,
  title={Multi-level quantum noise spectroscopy},
  author={Sung, Youngkyu and Veps{\"a}l{\"a}inen, Antti and Braum{\"u}ller, Jochen and Yan, Fei and Wang, Joel I-Jan and Kjaergaard, Morten and Winik, Roni and Krantz, Philip and Bengtsson, Andreas and Melville, Alexander J and others},
  journal={Nat. Commun.},
  volume={12},
  number={1},
  pages={967},
  year={2021},
  publisher={Nature Publishing Group UK London}
}

@article{ringbauer2022universal,
  title={A universal qudit quantum processor with trapped ions},
  author={Ringbauer, Martin and Meth, Michael and Postler, Lukas and Stricker, Roman and Blatt, Rainer and Schindler, Philipp and Monz, Thomas},
  journal={Nat. Phys.},
  volume={18},
  number={9},
  pages={1053--1057},
  year={2022},
  publisher={Nature Publishing Group UK London}
}

@article{zalivako2024towards,
AUTHOR = {Zalivako, Ilia V. and et al.},
TITLE = {Towards a Multiqudit Quantum Processor Based on a ${}^{171}${Yb}${}^+$ Ion String: Realizing Basic Quantum Algorithms},
JOURNAL = {Quantum Rep.},
VOLUME = {7},
YEAR = {2025},
NUMBER = {2},
ARTICLE-NUMBER = {19},
URL = {https://www.mdpi.com/2624-960X/7/2/19},
ISSN = {2624-960X},
DOI = {10.3390/quantum7020019}
}

@article{picard2025entanglement,
  title={Entanglement and iSWAP gate between molecular qubits},
  author={Picard, Lewis RB and Park, Annie J and Patenotte, Gabriel E and Gebretsadkan, Samuel and Wellnitz, David and Rey, Ana Maria and Ni, Kang-Kuen},
  journal={Nature},
  volume={637},
  number={8047},
  pages={821--826},
  year={2025},
  publisher={Nature Publishing Group UK London}
}

@article{fu2022experimental,
  title={Experimental investigation of quantum correlations in a two-qutrit spin system},
  author={Fu, Yue and Liu, Wenquan and Ye, Xiangyu and Wang, Ya and Zhang, Chengjie and Duan, Chang-Kui and Rong, Xing and Du, Jiangfeng},
  journal={Phys. Rev. Lett.},
  volume={129},
  number={10},
  pages={100501},
  year={2022},
  publisher={APS}
}

@article{
Anderegg2019optical,
author = {Loïc Anderegg  and Lawrence W. Cheuk  and Yicheng Bao  and Sean Burchesky  and Wolfgang Ketterle  and Kang-Kuen Ni  and John M. Doyle },
title = {An optical tweezer array of ultracold molecules},
journal = {Science},
volume = {365},
number = {6458},
pages = {1156-1158},
year = {2019},
doi = {10.1126/science.aax1265},
}

@article{will2016coherent,
  title={Coherent microwave control of ultracold Na 23 K 40 molecules},
  author={Will, Sebastian A and Park, Jee Woo and Yan, Zoe Z and Loh, Huanqian and Zwierlein, Martin W},
  journal={Phys. Rev. Lett.},
  volume={116},
  number={22},
  pages={225306},
  year={2016},
  publisher={APS}
}

@article{ji2020microwave,
  title={Microwave coherent control of ultracold ground-state molecules formed by short-range photoassociation},
  author={Ji, Zhonghua and Gong, Ting and He, Yonglin and Hutson, Jeremy M and Zhao, Yanting and Xiao, Liantuan and Jia, Suotang},
  journal={Physical Chemistry Chemical Physics},
  volume={22},
  number={23},
  pages={13002--13007},
  year={2020},
  publisher={Royal Society of Chemistry}
}

@article{park2017second,
  title={Second-scale nuclear spin coherence time of ultracold 23Na40K molecules},
  author={Park, Jee Woo and Yan, Zoe Z and Loh, Huanqian and Will, Sebastian A and Zwierlein, Martin W},
  journal={Science},
  volume={357},
  number={6349},
  pages={372--375},
  year={2017},
  publisher={American Association for the Advancement of Science}
}

@article{sawant2020ultracold,
  title={Ultracold polar molecules as qudits},
  author={Sawant, Rahul and Blackmore, Jacob A and Gregory, Philip D and Mur-Petit, Jordi and Jaksch, Dieter and Aldegunde, Jes{\'u}s and Hutson, Jeremy M and Tarbutt, Michael R and Cornish, Simon L},
  journal={New J. Phys.},
  volume={22},
  number={1},
  pages={013027},
  year={2020},
  publisher={IOP Publishing}
}

@article{shuman2010laser,
  title={Laser cooling of a diatomic molecule},
  author={Shuman, Edward S and Barry, John F and DeMille, David},
  journal={Nature},
  volume={467},
  number={7317},
  pages={820--823},
  year={2010},
  publisher={Nature Publishing Group UK London}
}

@article{ruttley2023formation,
  title={Formation of ultracold molecules by merging optical tweezers},
  author={Ruttley, Daniel K and Guttridge, Alexander and Spence, Stefan and Bird, Robert C and Le Sueur, C Ruth and Hutson, Jeremy M and Cornish, Simon L},
  journal={Phys. Rev. Lett.},
  volume={130},
  number={22},
  pages={223401},
  year={2023},
  publisher={APS}
}

@article{hughes2020robust,
  title={Robust entangling gate for polar molecules using magnetic and microwave fields},
  author={Hughes, Michael and Frye, Matthew D and Sawant, Rahul and Bhole, Gaurav and Jones, Jonathan A and Cornish, Simon L and Tarbutt, MR and Hutson, Jeremy M and Jaksch, Dieter and Mur-Petit, Jordi},
  journal={Phys. Rev. A},
  volume={101},
  number={6},
  pages={062308},
  year={2020},
  publisher={APS}
}

@article{herrera2014infrared,
  title={Infrared-dressed entanglement of cold open-shell polar molecules for universal matchgate quantum computing},
  author={Herrera, Felipe and Cao, Yudong and Kais, Sabre and Whaley, K Birgitta},
  journal={New J. Phys.},
  volume={16},
  number={7},
  pages={075001},
  year={2014},
  publisher={IOP Publishing}
}

@article{bause2020tune,
  title={Tune-out and magic wavelengths for ground-state Na 23 K 40 molecules},
  author={Bause, Roman and Li, Ming and Schindewolf, Andreas and Chen, Xing-Yan and Duda, Marcel and Kotochigova, Svetlana and Bloch, Immanuel and Luo, Xin-Yu},
  journal={Phys. Rev. Lett.},
  volume={125},
  number={2},
  pages={023201},
  year={2020},
  publisher={APS}
}

@article{blackmore2020controlling,
  title={Controlling the ac Stark effect of {RbCs} with dc electric and magnetic fields},
  author={Blackmore, Jacob A and Sawant, Rahul and Gregory, Philip D and Bromley, Sarah L and Aldegunde, Jes{\'u}s and Hutson, Jeremy M and Cornish, Simon L},
  journal={Phys. Rev. A},
  volume={102},
  number={5},
  pages={053316},
  year={2020},
  publisher={APS}
}

@article{burchesky2021rotational,
  title={Rotational coherence times of polar molecules in optical tweezers},
  author={Burchesky, Sean and Anderegg, Lo{\"\i}c and Bao, Yicheng and Yu, Scarlett S and Chae, Eunmi and Ketterle, Wolfgang and Ni, Kang-Kuen and Doyle, John M},
  journal={Phys. Rev. Lett.},
  volume={127},
  number={12},
  pages={123202},
  year={2021},
  publisher={APS}
}

@misc{grimm1999opticaldipoletrapsneutral,
      title={Optical dipole traps for neutral atoms}, 
      author={Rudolf Grimm and Matthias Weidemüller and Yurii B. Ovchinnikov},
      year={1999},
      eprint={physics/9902072},
      archivePrefix={arXiv},
      primaryClass={physics.atom-ph},
      url={https://arxiv.org/abs/physics/9902072}, 
}

@article{cline1994spin,
  title={Spin relaxation of optically trapped atoms by light scattering},
  author={Cline, RA and Miller, JD and Matthews, MR and Heinzen, DJ},
  journal={Optics letters},
  volume={19},
  number={3},
  pages={207--209},
  year={1994},
  publisher={Optical Society of America}
}

@article{caldwell2020sideband,
  title={Sideband cooling of molecules in optical traps},
  author={Caldwell, Luke and Tarbutt, MR},
  journal={Phys. Rev. Res.},
  volume={2},
  number={1},
  pages={013251},
  year={2020},
  publisher={APS}
}

@article{kotochigova2024rotational,
  title={Rotational magic conditions for ultracold molecules in the presence of Raman and Rayleigh scattering},
  author={Kotochigova, Svetlana and Guan, Qingze and Tiesinga, Eite and Scarola, Vito and DeMarco, Brian and Gadway, Bryce},
  journal={New J. Phys.},
  volume={26},
  number={6},
  pages={063025},
  year={2024},
  publisher={IOP Publishing}
}

@article{bergonzoni2025iswap,
  title={iSWAP gate with polar molecules: Robustness criteria for entangling operations},
  author={Bergonzoni, Matteo and Jandura, Sven and Pupillo, Guido},
  journal={Phys. Rev. A},
  volume={112},
  number={3},
  pages={032621},
  year={2025},
  publisher={APS}
}

@article{haimberger2009formation,
  title={Formation of ultracold, highly polar {${X}\,{}^1{\Sigma}^+$} {NaCs} molecules},
  author={Haimberger, C and Kleinert, J and Zabawa, P and Wakim, A and Bigelow, NP},
  journal={New J. Phys.},
  volume={11},
  number={5},
  pages={055042},
  year={2009},
  publisher={IOP Publishing}
}

@article{florshaim2024spatial,
  title={Spatial adiabatic passage of ultracold atoms in optical tweezers},
  author={Florshaim, Yanay and Zohar, Elad and Koplovich, David Zeev and Meltzer, Ilan and Weill, Rafi and Nemirovsky, Jonathan and Stern, Amir and Sagi, Yoav},
  journal={Science Advances},
  volume={10},
  number={40},
  pages={eadl1220},
  year={2024},
  publisher={American Association for the Advancement of Science}
}

@misc{sWill,
  author={S. Will},
  note = {private communication}
}

@article{carr2009cold,
  title={Cold and ultracold molecules: science, technology and applications},
  author={Carr, Lincoln D and DeMille, David and Krems, Roman V and Ye, Jun},
  journal={New J. Phys.},
  volume={11},
  number={5},
  pages={055049},
  year={2009},
  publisher={IOP Publishing}
}

@article{mitra2022quantum,
  title={Quantum control of molecules for fundamental physics},
  author={Mitra, D and Leung, KH and Zelevinsky, T},
  journal={Phys. Rev. A},
  volume={105},
  number={4},
  pages={040101},
  year={2022},
  publisher={APS}
}

@article{hudson2006cold,
  title={Cold molecule spectroscopy for constraining the evolution of the fine structure constant},
  author={Hudson, Eric R and Lewandowski, HJ and Sawyer, Brian C and Ye, Jun},
  journal={Phys. Rev. Lett.},
  volume={96},
  number={14},
  pages={143004},
  year={2006},
  publisher={APS}
}

@article{zelevinsky2008precision,
  title={Precision test of mass-ratio variations with lattice-confined ultracold molecules},
  author={Zelevinsky, T and Kotochigova, S and Ye, Jun},
  journal={Phys. Rev. Lett.},
  volume={100},
  number={4},
  pages={043201},
  year={2008},
  publisher={APS}
}

@article{demille2008enhanced,
  title={Enhanced sensitivity to variation of me/mp in molecular spectra},
  author={DeMille, D and Sainis, S and Sage, J and Bergeman, T and Kotochigova, S and Tiesinga, E},
  journal={Phys. Rev. Lett.},
  volume={100},
  number={4},
  pages={043202},
  year={2008},
  publisher={APS}
}

@inproceedings{sauer2006probing,
  title={Probing the electron EDM with cold molecules},
  author={Sauer, BE and Ashworth, HT and Hudson, JJ and Tarbutt, MR and Hinds, EA},
  booktitle={AIP conference proceedings},
  volume={869},
  number={1},
  pages={44--51},
  year={2006},
  organization={American Institute of Physics}
}

@inproceedings{vutha2008search,
  title={Search for the electron's electric dipole moment with a cold molecular beam of ThO},
  author={Vutha, Amar C and Baker, O Keith and Campbell, Wesley C and DeMille, David and Doyle, John M and Gabrielse, Gerald and Gurevich, Yulia V and Jansen, Maarten AHM},
  booktitle={APS Division of Atomic, Molecular and Optical Physics Meeting Abstracts},
  volume={39},
  pages={OPC--50},
  year={2008}
}

@article{gregory2024second,
  title={Second-scale rotational coherence and dipolar interactions in a gas of ultracold polar molecules},
  author={Gregory, Philip D and Fernley, Luke M and Tao, Albert Li and Bromley, Sarah L and Stepp, Jonathan and Zhang, Zewen and Kotochigova, Svetlana and Hazzard, Kaden RA and Cornish, Simon L},
  journal={Nat. Phys.},
  volume={20},
  number={3},
  pages={415--421},
  year={2024},
  publisher={Nature Publishing Group UK London}
}

@article{gregory2021robust,
  title={Robust storage qubits in ultracold polar molecules},
  author={Gregory, Philip D and Blackmore, Jacob A and Bromley, Sarah L and Hutson, Jeremy M and Cornish, Simon L},
  journal={Nat. Phys.},
  volume={17},
  number={10},
  pages={1149--1153},
  year={2021},
  publisher={Nature Publishing Group UK London}
}

@article{zhang2022creation,
  title={Creation of high-dimensional entanglement of polar molecules via optimal control fields},
  author={Zhang, Zuo-Yuan and Liu, Jin-Ming},
  journal={Phys. Rev. A},
  volume={105},
  number={2},
  pages={023113},
  year={2022},
  publisher={APS}
}

@article{andre2006coherent,
  title={A coherent all-electrical interface between polar molecules and mesoscopic superconducting resonators},
  author={Andr{\'e}, Axel and DeMille, David and Doyle, John M and Lukin, Mikhail D and Maxwell, St Ex and Rabl, Peter and Schoelkopf, Robert J and Zoller, Peter},
  journal={Nat. Phys.},
  volume={2},
  number={9},
  pages={636--642},
  year={2006},
  publisher={Nature Publishing Group UK London}
}

@article{ospelkaus2010controlling,
  title={Controlling the hyperfine state of rovibronic ground-state polar molecules},
  author={Ospelkaus, S and Ni, K-K and Qu{\'e}m{\'e}ner, G and Neyenhuis, B and Wang, D and De Miranda, MHG and Bohn, JL and Ye, J and Jin, DS},
  journal={Phys. Rev. Lett.},
  volume={104},
  number={3},
  pages={030402},
  year={2010},
  publisher={APS}
}

@article{yan2013observation,
  title={Observation of dipolar spin-exchange interactions with lattice-confined polar molecules},
  author={Yan, Bo and Moses, Steven A and Gadway, Bryce and Covey, Jacob P and Hazzard, Kaden RA and Rey, Ana Maria and Jin, Deborah S and Ye, Jun},
  journal={Nature},
  volume={501},
  number={7468},
  pages={521--525},
  year={2013},
  publisher={Nature Publishing Group UK London}
}

@article{hazzard2014many,
  title={Many-body dynamics of dipolar molecules in an optical lattice},
  author={Hazzard, Kaden RA and Gadway, Bryce and Foss-Feig, Michael and Yan, Bo and Moses, Steven A and Covey, Jacob P and Yao, Norman Y and Lukin, Mikhail D and Ye, Jun and Jin, Deborah S and others},
  journal={Phys. Rev. Lett.},
  volume={113},
  number={19},
  pages={195302},
  year={2014},
  publisher={APS}
}

@article{kaufman2021quantum,
  title={Quantum science with optical tweezer arrays of ultracold atoms and molecules},
  author={Kaufman, Adam M and Ni, Kang-Kuen},
  journal={Nat. Phys.},
  volume={17},
  number={12},
  pages={1324--1333},
  year={2021},
  publisher={Nature Publishing Group UK London}
}

@article{yelin2006schemes,
  title={Schemes for robust quantum computation with polar molecules},
  author={Yelin, SF and Kirby, K and C{\^o}t{\'e}, Robin},
  journal={Phys. Rev. A},
  volume={74},
  number={5},
  pages={050301},
  year={2006},
  publisher={APS}
}

@article{wei2011entanglement,
  title={Entanglement of polar symmetric top molecules as candidate qubits},
  author={Wei, Qi and Kais, Sabre and Friedrich, Bretislav and Herschbach, Dudley},
  journal={The Journal of chemical physics},
  volume={135},
  number={15},
  year={2011},
  publisher={AIP Publishing}
}

@article{ni2018dipolar,
  title={Dipolar exchange quantum logic gate with polar molecules},
  author={Ni, Kang-Kuen and Rosenband, Till and Grimes, David D},
  journal={Chemical science},
  volume={9},
  number={33},
  pages={6830--6838},
  year={2018},
  publisher={Royal Society of Chemistry}
}

@article{bao2023dipolar,
  title={Dipolar spin-exchange and entanglement between molecules in an optical tweezer array},
  author={Bao, Yicheng and Yu, Scarlett S and Anderegg, Lo{\"\i}c and Chae, Eunmi and Ketterle, Wolfgang and Ni, Kang-Kuen and Doyle, John M},
  journal={Science},
  volume={382},
  number={6675},
  pages={1138--1143},
  year={2023},
  publisher={American Association for the Advancement of Science}
}

@article{zhang2022optical,
  title={An optical tweezer array of ground-state polar molecules},
  author={Zhang, Jessie T and Picard, Lewis RB and Cairncross, William B and Wang, Kenneth and Yu, Yichao and Fang, Fang and Ni, Kang-Kuen},
  journal={Quantum Science and Technology},
  volume={7},
  number={3},
  pages={035006},
  year={2022},
  publisher={IOP Publishing}
}

@article{ruttley2024enhanced,
  title={Enhanced quantum control of individual ultracold molecules using optical tweezer arrays},
  author={Ruttley, Daniel K and Guttridge, Alexander and Hepworth, Tom R and Cornish, Simon L},
  journal={PRX quantum},
  volume={5},
  number={2},
  pages={020333},
  year={2024},
  publisher={APS}
}

@article{caldwell2020long,
  title={Long rotational coherence times of molecules in a magnetic trap},
  author={Caldwell, L and Williams, HJ and Fitch, NJ and Aldegunde, J and Hutson, Jeremy M and Sauer, BE and Tarbutt, MR},
  journal={Phys. Rev. Lett.},
  volume={124},
  number={6},
  pages={063001},
  year={2020},
  publisher={APS}
}

@article{blackmore2018ultracold,
  title={Ultracold molecules for quantum simulation: rotational coherences in {CaF} and {RbCs}},
  author={Blackmore, Jacob A and Caldwell, Luke and Gregory, Philip D and Bridge, Elizabeth M and Sawant, Rahul and Aldegunde, Jes{\'u}s and Mur-Petit, Jordi and Jaksch, Dieter and Hutson, Jeremy M and Sauer, BE and others},
  journal={Quantum Science and Technology},
  volume={4},
  number={1},
  pages={014010},
  year={2018},
  publisher={IOP Publishing}
}

@article{micheli2006toolbox,
  title={A toolbox for lattice-spin models with polar molecules},
  author={Micheli, Andrea and Brennen, Gavin K and Zoller, Peter},
  journal={Nat. Phys.},
  volume={2},
  number={5},
  pages={341--347},
  year={2006},
  publisher={Nature Publishing Group UK London}
}

@article{gorshkov2011tunable,
  title={Tunable superfluidity and quantum magnetism with ultracold polar molecules},
  author={Gorshkov, Alexey V and Manmana, Salvatore R and Chen, Gang and Ye, Jun and Demler, Eugene and Lukin, Mikhail D and Rey, Ana Maria},
  journal={Phys. Rev. Lett.},
  volume={107},
  number={11},
  pages={115301},
  year={2011},
  publisher={APS}
}

@article{hepworth2025long,
  title={Long-lived multilevel coherences and spin-1 dynamics encoded in the rotational states of ultracold molecules},
  author={Hepworth, Tom R and Ruttley, Daniel K and von Gierke, Fritz and Gregory, Philip D and Guttridge, Alexander and Cornish, Simon L},
  journal={Nat. Commun.},
  volume={16},
  number={1},
  pages={7131},
  year={2025},
  publisher={Nature Publishing Group UK London}
}

@article{muthukrishnan2000multivalued,
  title={Multivalued logic gates for quantum computation},
  author={Muthukrishnan, Ashok and Stroud Jr, Carlos R},
  journal={Phys. Rev. A},
  volume={62},
  number={5},
  pages={052309},
  year={2000},
  publisher={APS}
}

@article{Bartlett2002Quantum,
  title = {Quantum encodings in spin systems and harmonic oscillators},
  author = {Bartlett, Stephen D. and de Guise, Hubert and Sanders, Barry C.},
  journal = {Phys. Rev. A},
  volume = {65},
  issue = {5},
  pages = {052316},
  numpages = {4},
  year = {2002},
  month = {May},
  publisher = {American Physical Society},
  doi = {10.1103/PhysRevA.65.052316}
}

@article{zilic2007scaling,
  title={Scaling and better approximating quantum Fourier transform by higher radices},
  author={Zilic, Zeljko and Radecka, Katarzyna},
  journal={IEEE Transactions on computers},
  volume={56},
  number={2},
  pages={202--207},
  year={2007},
  publisher={IEEE}
}

@INPROCEEDINGS{Parasa2011Quantum,
  author={Parasa, Vamsi and Perkowski, Marek},
  booktitle={2011 41st IEEE International Symposium on Multiple-Valued Logic}, 
  title={Quantum Phase Estimation Using Multivalued Logic}, 
  year={2011},
  volume={},
  number={},
  pages={224-229},
  doi={10.1109/ISMVL.2011.47}}

@article{campbell2012magic,
  title={Magic-state distillation in all prime dimensions using quantum reed-muller codes},
  author={Campbell, Earl T and Anwar, Hussain and Browne, Dan E},
  journal={Phys. Rev. X},
  volume={2},
  number={4},
  pages={041021},
  year={2012},
  publisher={APS}
}

@article{duclos2013kitaev,
  title={Kitaev's Z d-code threshold estimates},
  author={Duclos-Cianci, Guillaume and Poulin, David},
  journal={Phys. Rev. A},
  volume={87},
  number={6},
  pages={062338},
  year={2013},
  publisher={APS}
}

@article{anwar2014fast,
  title={Fast decoders for qudit topological codes},
  author={Anwar, Hussain and Brown, Benjamin J and Campbell, Earl T and Browne, Dan E},
  journal={New J. Phys.},
  volume={16},
  number={6},
  pages={063038},
  year={2014},
  publisher={IOP Publishing}
}

@article{campbell2014enhanced,
  title={Enhanced fault-tolerant quantum computing in d-level systems},
  author={Campbell, Earl T},
  journal={Phys. Rev. Lett.},
  volume={113},
  number={23},
  pages={230501},
  year={2014},
  publisher={APS}
}

@article{andrist2015error,
  title={Error thresholds for Abelian quantum double models: Increasing the bit-flip stability of topological quantum memory},
  author={Andrist, Ruben S and Wootton, James R and Katzgraber, Helmut G},
  journal={Phys. Rev. A},
  volume={91},
  number={4},
  pages={042331},
  year={2015},
  publisher={APS}
}

@article{caldwell2020enhancing,
  title={Enhancing dipolar interactions between molecules using state-dependent optical tweezer traps},
  author={Caldwell, L and Tarbutt, MR},
  journal={Phys. Rev. Lett.},
  volume={125},
  number={24},
  pages={243201},
  year={2020},
  publisher={APS}
}

@article{caldwell2021general,
  title={General approach to state-dependent optical-tweezer traps for polar molecules},
  author={Caldwell, L and Tarbutt, MR},
  journal={Phys. Rev. Res.},
  volume={3},
  number={1},
  pages={013291},
  year={2021},
  publisher={APS}
}

@article{ruttley2025long,
  title={Long-lived entanglement of molecules in magic-wavelength optical tweezers},
  author={Ruttley, Daniel K and Hepworth, Tom R and Guttridge, Alexander and Cornish, Simon L},
  journal={Nature},
  volume={637},
  number={8047},
  pages={827--832},
  year={2025},
  publisher={Nature Publishing Group UK London}
}

@article{moses2015creation,
  title={Creation of a low-entropy quantum gas of polar molecules in an optical lattice},
  author={Moses, Steven A and Covey, Jacob P and Miecnikowski, Matthew T and Yan, Bo and Gadway, Bryce and Ye, Jun and Jin, Deborah S},
  journal={Science},
  volume={350},
  number={6261},
  pages={659--662},
  year={2015},
  publisher={American Association for the Advancement of Science}
}

@article{wall2015realizing,
  title={Realizing unconventional quantum magnetism with symmetric top molecules},
  author={Wall, ML and Maeda, Kenji and Carr, Lincoln D},
  journal={New J. Phys.},
  volume={17},
  number={2},
  pages={025001},
  year={2015},
  publisher={IOP Publishing}
}

@article{manmana2012topological,
  title={Topological phases in ultracold polar-molecule quantum magnets},
  author={Manmana, Salvatore R and Stoudenmire, EM and Hazzard, Kaden RA and Rey, Ana Maria and Gorshkov, Alexey V},
  journal={arXiv preprint arXiv:1210.5518},
  year={2012}
}

@article{anderegg2018laser,
  title={Laser cooling of optically trapped molecules},
  author={Anderegg, Lo{\"\i}c and Augenbraun, Benjamin L and Bao, Yicheng and Burchesky, Sean and Cheuk, Lawrence W and Ketterle, Wolfgang and Doyle, John M},
  journal={Nat. Phys.},
  volume={14},
  number={9},
  pages={890--893},
  year={2018},
  publisher={Nature Publishing Group UK London}
}

@article{watson2015qudit,
  title={Qudit color codes and gauge color codes in all spatial dimensions},
  author={Watson, Fern HE and Campbell, Earl T and Anwar, Hussain and Browne, Dan E},
  journal={Phys. Rev. A},
  volume={92},
  number={2},
  pages={022312},
  year={2015},
  publisher={APS}
}

@article{krishna2019towards,
  title={Towards low overhead magic state distillation},
  author={Krishna, Anirudh and Tillich, Jean-Pierre},
  journal={Phys. Rev. Lett.},
  volume={123},
  number={7},
  pages={070507},
  year={2019},
  publisher={APS}
}

@article{ni2008high,
  title={A high phase-space-density gas of polar molecules},
  author={Ni, K-K and Ospelkaus, S and De Miranda, MHG and Pe'Er, A and Neyenhuis, B and Zirbel, JJ and Kotochigova, S and Julienne, PS and Jin, DS and Ye, Jun},
  journal={Science},
  volume={322},
  number={5899},
  pages={231--235},
  year={2008},
  publisher={American Association for the Advancement of Science}
}

@article{ospelkaus2010quantum,
  title={Quantum-state controlled chemical reactions of ultracold potassium-rubidium molecules},
  author={Ospelkaus, S and Ni, K-K and Wang, D and De Miranda, MHG and Neyenhuis, B and Qu{\'e}m{\'e}ner, G and Julienne, PS and Bohn, JL and Jin, DS and Ye, J},
  journal={Science},
  volume={327},
  number={5967},
  pages={853--857},
  year={2010},
  publisher={American Association for the Advancement of Science}
}

@article{deiglmayr2008formation,
  title={Formation of ultracold polar molecules in the rovibrational ground state},
  author={Deiglmayr, J and Grochola, A and Repp, M and M{\"o}rtlbauer, K and Gl{\"u}ck, C and Lange, J and Dulieu, O and Wester, R and Weidem{\"u}ller, M},
  journal={Phys. Rev. Lett.},
  volume={101},
  number={13},
  pages={133004},
  year={2008},
  publisher={APS}
}

@article{ni2010dipolar,
  title={Dipolar collisions of polar molecules in the quantum regime},
  author={Ni, K-K and Ospelkaus, S and Wang, D and Qu{\'e}m{\'e}ner, G and Neyenhuis, Brian and De Miranda, MHG and Bohn, JL and Ye, Jun and Jin, DS},
  journal={Nature},
  volume={464},
  number={7293},
  pages={1324--1328},
  year={2010},
  publisher={Nature Publishing Group UK London}
}

@article{de2011controlling,
  title={Controlling the quantum stereodynamics of ultracold bimolecular reactions},
  author={De Miranda, MHG and Chotia, A and Neyenhuis, Brian and Wang, D and Qu{\'e}m{\'e}ner, G and Ospelkaus, Silke and Bohn, JL and Ye, Jun and Jin, DS},
  journal={Nat. Phys.},
  volume={7},
  number={6},
  pages={502--507},
  year={2011},
  publisher={Nature Publishing Group UK London}
}

@misc{SM,
  note = {See Supplemental Material at [\url{https://journals.aps.org/prresearch/supplemental/10.1103/tcqw-ll6v/supplementary.pdf}] for detailed derivations and numerical analysis, which includes Refs.~\cite{aldegunde2017hyperfine,bonczyk1967hyperfine,PhysRevA.97.042505,childs1981radio,caldwell2020long,park2017second,lin2022seconds}.}
}

@article{zhou2024robust,
  title={Robust Hamiltonian engineering for interacting qudit systems},
  author={Zhou, Hengyun and Gao, Haoyang and Leitao, Nathaniel T and Makarova, Oksana and Cong, Iris and Douglas, Alexander M and Martin, Leigh S and Lukin, Mikhail D},
  journal={Phys. Rev. X},
  volume={14},
  number={3},
  pages={031017},
  year={2024},
  publisher={APS}
}

@article{nikolaeva2024efficient,
  title={Efficient realization of quantum algorithms with qudits},
  author={Nikolaeva, Anastasiia S and Kiktenko, Evgeniy O and Fedorov, Aleksey K},
  journal={EPJ Quantum Technology},
  volume={11},
  number={1},
  pages={1--25},
  year={2024},
  publisher={Springer}
}

@article{nikolaeva2022decomposing,
  title={Decomposing the generalized Toffoli gate with qutrits},
  author={Nikolaeva, Anastasiia S and Kiktenko, Evgeniy O and Fedorov, Aleksey K},
  journal={Phys. Rev. A},
  volume={105},
  number={3},
  pages={032621},
  year={2022},
  publisher={APS}
}

@article{williams2011quantum,
  title={Quantum gates},
  author={Williams, Colin P and Williams, Colin P},
  journal={Explorations in quantum computing},
  pages={51--122},
  year={2011},
  publisher={Springer}
}

@BOOK{landau2013quantum,
  title={Quantum mechanics: non-relativistic theory},
  author={Landau, Lev Davidovich and Lifshitz, Evgenii Mikhailovich},
  volume={3},
  year={2013},
  publisher={Elsevier}
}

@ARTICLE{ernst1985electric,
  title={Electric dipole moment of {SrF} {${X}\,{}^2{\Sigma}^+$} from high-precision stark effect measurements},
  author={Ernst, WE and K{\"a}ndler, J{\"o}rn and Kindt, S and T{\"o}rring, T},
  journal={Chemical physics letters},
  volume={113},
  number={4},
  pages={351--354},
  year={1985},
  publisher={Elsevier}
}

@ARTICLE{hao2019high,
  title={High accuracy theoretical investigations of {CaF}, {SrF}, and {BaF} and implications for laser-cooling},
  author={Hao, Yongliang and Pa{\v{s}}teka, Luk{\'a}{\v{s}} F and Visscher, Lucas and Aggarwal, Parul and Bethlem, Hendrick L and Boeschoten, Alexander and Borschevsky, Anastasia and Denis, Malika and Esajas, Kevin and Hoekstra, Steven and others},
  journal={The Journal of chemical physics},
  volume={151},
  number={3},
  year={2019},
  publisher={AIP Publishing}
}

\end{document}